\documentclass[10pt]{article}

\usepackage[a4paper]{geometry}
\usepackage[english]{babel}
\usepackage{array}
\usepackage{amsmath,amsthm,amsfonts,amssymb}
\usepackage{unicode}
\usepackage{subcaption}
\usepackage[type={CC},modifier={by-nc},imagemodifier={-eu},version={4.0},imagewidth=5em]{doclicense}
\usepackage{fancyhdr}

\usepackage{algorithmic}

\algsetup{linenodelimiter=.}

\usepackage[pdfusetitle]{hyperref}
\hypersetup{
  unicode=true,
  colorlinks=true,
  citecolor=blue!70!black,
  filecolor=black,
  linkcolor=red!70!black,
  urlcolor=blue,
  pdfstartview={FitH},
  pdfauthor={Luca De Feo},
  pdfsubject={Mathematics},
  pdfkeywords={Cryptography, Number theory, Elliptic curves, Isogenies},
}

\usepackage{tikz}
\usetikzlibrary{arrows,matrix,decorations,decorations.text,decorations.pathmorphing,calc}
\pgfkeys{/triangle/.code=\tikzset{x={(-0.5cm,-0.866cm)},y={(1cm,0cm)}}}
\pgfkeys{/lattice/.code n args={4}{\tikzset{cm={#1,#2,#3,#4,(0,0)}}}}

\newcommand{\axes}[4]{
  \clip (#1,#3) rectangle (#2,#4);
  \draw [thin, gray, -latex] (#1,0) -- (#2,0);
  \draw [thin, gray, -latex] (0,#3) -- (0,#4);
}

\newcommand{\lattice}[3][2pt]{
  \draw[style=help lines,dashed] (#2-1,#2-1) grid[step=1] (#3+1,#3+1);
  \foreach \x in {#2,...,#3}{
    \foreach \y in {#2,...,#3}{
      \node[draw,circle,inner sep=#1,fill] at (\x,\y) {};
    }
  }
}

\theoremstyle{plain}
\newtheorem{theorem}{Theorem}

\newtheorem{corollary}[theorem]{Corollary}
\newtheorem{proposition}[theorem]{Proposition}
\theoremstyle{definition}
\newtheorem{definition}[theorem]{Definition}

\newtheorem{problem}{Problem}
\newtheorem{exercice}{Exercice}[part]

\DeclareMathOperator{\End}{End} 
\DeclareMathOperator{\loglog}{loglog}
\DeclareMathOperator{\im}{Im}
\DeclareMathOperator{\GL}{GL}
\DeclareMathOperator{\SL}{SL}
\DeclareMathOperator{\Cl}{Cl}
\DeclareMathOperator{\Ell}{Ell}

\def\P{\ensuremath{\mathbb{P}}}
\def\F{\ensuremath{\mathbb{F}}}
\def\O{\ensuremath{\mathcal{O}}}
\def\tildO{\ensuremath{\tilde{O}}}

\def\a{\ensuremath{\mathfrak{a}}}

\newcommand{\bl}[1]{\textcolor{blue}{#1}}
\newcommand{\rd}[1]{\textcolor{red}{#1}}

\title{Mathematics of Isogeny Based Cryptography}
\author{Luca De Feo\\
  Universit\'e de Versailles \& Inria Saclay\\
  \url{http://defeo.lu/}}
\date{\'Ecole math\'ematique africaine\\
  May 10 -- 23, 2017, Thi\`es, Senegal}

\begin{document}
\maketitle
\thispagestyle{fancy}
\renewcommand{\headrulewidth}{0pt}
\renewcommand{\footrulewidth}{0.4pt}
\cfoot{\doclicenseThis}
\lfoot{\LaTeX{} source code available at \url{https://github.com/defeo/ema2017/}.}

\section*{Introduction}

These lectures notes were written for a summer school on
\emph{Mathematics for post-quantum cryptography} in Thiès, Senegal. %
They try to provide a guide for Masters' students to get through the
vast literature on elliptic curves, without getting lost on their way
to learning isogeny based cryptography. %
They are by no means a reference text on the theory of elliptic
curves, nor on cryptography; students are encouraged to complement
these notes with some of the books recommended in the bibliography. %

The presentation is divided in three parts, roughly corresponding to
the three lectures given. %
In an effort to keep the reader interested, each part alternates
between the fundamental theory of elliptic curves, and applications in
cryptography. %
We often prefer to have the main ideas flow smoothly, rather than
having a rigorous presentation as one would have in a more classical
book. %
The reader will excuse us for the inaccuracies and the omissions.

\paragraph{Isogeny Based Cryptography} is a very young field, that has
only begun in the 2000s. %
It has its roots in \emph{Elliptic Curve Cryptography} (ECC), a
somewhat older branch of public-key cryptography that was started in
the 1980s, when Miller and Koblitz first suggested to use elliptic
curves inside the Diffie-Hellman key exchange protocol (see
Section~\ref{sec:appl-diff-hellm}). %

ECC only started to gain traction in the 1990s, after Schoof's
algorithm made it possible to easily find elliptic curves of large
prime order. %
It is nowadays a staple in public-key cryptography. %
The 2000s have seen two major innovations in ECC: the rise of
\emph{Pairing Based Cryptography} (PBC), epitomized by Joux' one-round
tripartite Diffie-Hellman key exchange, and the advent of
Isogeny based cryptography, initiated by the works of Couveignes,
Teske and Rostovtsev \& Stolbunov. %
While PBC has attracted most of the attention during the first decade,
thanks to its revolutionary applications, isogeny based cryptography
has stayed mostly discrete during this time. %
It is only in the second half of the 2010 that the attention has
partly shifted to isogenies. %
The main reason for this is the sudden realization by the
cryptographic community of the very possibly near arrival of a
\emph{general purpose quantum computer}. %
While the capabilities of such futuristic machine would render all of ECC
and PBC suddenly worthless, isogeny based cryptography seems to resist
much better to the cryptanalytic powers of the quantum computer.

In these notes, after a review of the general theory of elliptic
curves and isogenies, we will present the most important isogeny based
systems, and their cryptographic properties.

{
  \hypersetup{linkcolor=black}
  \setcounter{tocdepth}{1}
  \tableofcontents
}


\clearpage
\part{Elliptic curves and cryptography}

Throughout this part we let $k$ be a field, and we denote by
$\bar{k}$ its algebraic closure. %
We review the basic theory of elliptic curves, and two classic
applications in cryptography. %
The interested reader will find more details on elliptic curves
in~\cite{silverman:elliptic}, and on their use in cryptography
in~\cite{joux2009algorithmic,galbraith2012mathematics}.

\section{Elliptic curves}

Elliptic curves are projective curves of genus 1 having a specified
base point. %
Projective space initially appeared through the process of adding
\emph{points at infinity}, as a method to understand the geometry of
projections (also known as \emph{perspective} in classical
painting). %
In modern terms, we define projective space as the collection of all
lines in affine space passing through the origin.

\begin{definition}[Projective space]
  The \emph{projective space of dimension $n$}, denoted by $\P^n$ or
  $\P^n(\bar{k})$, is the set of all $(n+1)$-tuples
  \[(x_0,\dots,x_n) ∈ \bar{k}^{n+1}\] %
  such that $(x_0,\dots,x_n) ≠ (0,\dots,0)$, taken modulo the
  equivalence relation
  \[(x_0,\dots,x_n) \sim (y_0,\dots,y_n)\] %
  if and only if there exists $λ\in\bar{k}$ such that
  $x_i=λ_iy_i$ for all $i$.
\end{definition}

The equivalence class of a projective point $(x_0,\dots,x_n)$ is
customarily denoted by $(x_0:\cdots:x_n)$. %
The set of the \emph{$k$-rational points}, denoted by $\P^n(k)$, is
defined as
\[\P^n(k) = \left\{(x_0:\cdots:x_n)∈\P^n\;\middle|\; x_i ∈ k \text{ for all $i$}\right\}.\] %
By fixing arbitrarily the coordinate $x_n=0$, we define a projective
space of dimension $n-1$, which we call the \emph{space at infinity};
its points are called \emph{points at infinity}.

From now on we suppose that the field $k$ has characteristic different
from $2$ and $3$. %
This has the merit of greatly simplifying the representation of an
elliptic curve. %
For a general definition, see~\cite[Chap.~III]{silverman:elliptic}.

\begin{definition}[Weierstrass equation]
  An \emph{elliptic curve} defined over $k$ is the locus in
  $\P^2(\bar{k})$ of an equation
  \begin{equation}
    \label{eq:weierstrass}
    Y^2Z = X^3 + aXZ^2 + bZ^3,    
  \end{equation}
  with $a,b∈k$ and $4a^3+27b^2\ne0$.

  The point $(0:1:0)$ is the only point on the line $Z=0$; it is
  called the \emph{point at infinity} of the curve.
\end{definition}

It is customary to write Eq.~\eqref{eq:weierstrass} in \emph{affine
  form}. %
By defining the coordinates $x=X/Z$ and $y=Y/Z$, we equivalently
define the elliptic curve as the locus of the equation
\[y^2 = x^3 + ax +b,\]
plus the point at infinity $\O=(0:1:0)$.

In characteristic different from $2$ and $3$, we can show that any
projective curve of genus $1$ with a distinguished point $\O$ is
isomorphic to a Weierstrass equation by sending $\O$ onto the point at
infinity $(0:1:0)$.

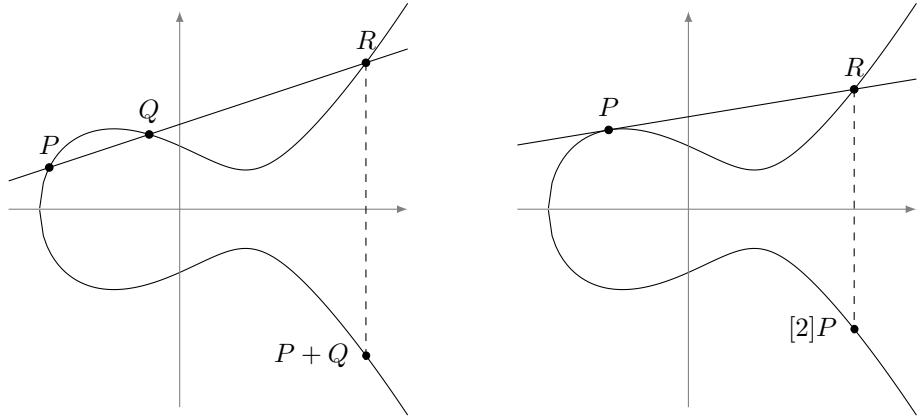
\begin{figure}
  \centering
  \hfill
  \begin{tikzpicture}[domain=-2.4566:4,samples=100,yscale=3/8,xscale=3/4]
    \draw plot (\x,{sqrt(\x*\x*\x-4*\x+5)});
    \draw plot (\x,{-sqrt(\x*\x*\x-4*\x+5)});

    \draw[thin,gray,-latex] (0,-7) -- (0,7);
    \draw[thin,gray,-latex] (-3,0) -- (4,0);

    \draw (-3,1) -- (4,8/3+3);
    \begin{scope}[every node/.style={draw,circle,inner sep=1pt,fill},cm={1,2/3,0,0,(0,3)}]
      \node at (-2.287980,0) {};
      \node at (-0.535051,0) {};
      \node at (3.267475,0) {};
    \end{scope}
    \begin{scope}[every node/.style={yshift=0.3cm},cm={1,2/3,0,0,(0,3)}]
      \node at (-2.287980,0) {$P$};
      \node at (-0.535051,0) {$Q$};
      \node at (3.267475,0) {$R$};
    \end{scope}

    \draw[dashed] (3.267475,3.267475*2/3+3) -- (3.267475,-3.267475*2/3-3) 
    node[draw,circle,inner sep=1pt,fill] {}
    node[xshift=-0.1cm,anchor=east] {$P+Q$};
  \end{tikzpicture}
  \hfill
  \begin{tikzpicture}[domain=-2.4566:4,samples=100,yscale=3/8,xscale=3/4]
    \draw plot (\x,{sqrt(\x*\x*\x-4*\x+5)});
    \draw plot (\x,{-sqrt(\x*\x*\x-4*\x+5)});

    \draw[thin,gray,-latex] (0,-7) -- (0,7);
    \draw[thin,gray,-latex] (-3,0) -- (4,0);
    
    \def\c{3.269524}
    \def\P{-1.398674}
    \def\R{2.908459}
    \draw (-3,-1+\c) -- (4,4/3+\c);
    \begin{scope}[every node/.style={draw,circle,inner sep=1pt,fill},cm={1,1/3,0,0,(0,3.269524)}]
      \node at (\P,0) {};
      \node at (\R,0) {};
    \end{scope}
    \begin{scope}[every node/.style={yshift=0.3cm},cm={1,1/3,0,0,(0,3.269524)}]
      \node at (\P,0) {$P$};
      \node at (\R,0) {$R$};
    \end{scope}

    \draw[dashed] (\R,\R/3+\c) -- (\R,-\R/3-\c) 
    node[draw,circle,inner sep=1pt,fill] {}
    node[xshift=-0.1cm,anchor=east] {$[2]P$};
  \end{tikzpicture}
  \hfill
  \strut
  
  \caption{An elliptic curve defined over $ℝ$, and the geometric
    representation of its group law.}
  \label{fig:weierstrass}
\end{figure}

Now, since any elliptic curve is defined by a cubic equation, Bezout's
theorem tells us that any line in $\P^2$ intersects the curve in
exactly three points, taken with multiplicity. %
We define a group law by requiring that three co-linear points sum to
zero. %

\begin{definition}
  Let $E\;:\;y^2=x^3+ax+b$ be an elliptic curve. Let $P_1=(x_1,y_1)$
  and $P_2=(x_2,y_2)$ be two points on $E$ different from the point at
  infinity, then we define a composition law $⊕$ on $E$ as
  follows:
  \begin{itemize}
  \item $P ⊕ \O = \O ⊕ P = P$ for any point $P∈E$;
  \item If $x_1=x_2$ and $y_1=-y_2$, then $P_1⊕P_2 = \O$;
  \item Otherwise set
    \[λ =
      \begin{cases}
        \frac{y_2-y_1}{x_2-x_1} &\text{if $P≠Q$,}\\
        \frac{3x_1^2+a}{2y_1} &\text{if $P=Q$,}
      \end{cases}
    \]
    then the point $(P_1⊕P_2)=(x_3,y_3)$ is defined by
    \begin{align*}
      x_3 &= λ^2 - x_1 - x_2,\\
      y_3 &= -λx_3 - y_1 + λx_1.
    \end{align*}
  \end{itemize}
\end{definition}

It can be shown that the above law defines an Abelian group, thus we
will simply write $+$ for $⊕$. %
The $n$-th scalar multiple of a point $P$ will be denoted by $[n]P$. %
When $E$ is defined over $k$, the subgroup of its \emph{rational
  points over $k$} is customarily denoted $E(k)$. %
Figure~\ref{fig:weierstrass} shows a graphical depiction of the group
law on an elliptic curve defined over $ℝ$.

We now turn to the group structure of elliptic curves. %
The torsion part is easily characterized.

\begin{proposition}
  Let $E$ be an elliptic curve defined over a field $k$, and let $m≠0$
  be an integer. %
  The $m$-torsion group of $E$, denoted by $E[m]$, has the following
  structure:
  \begin{itemize}
  \item $E[m] ≃ (ℤ/mℤ)^2$ if the characteristic of $k$ does not divide
    $m$;
  \item If $p>0$ is the characteristic of $k$, then 
    \[E[p^i] ≃
      \begin{cases}
        ℤ/p^iℤ & \text{for any $i≥0$, or}\\
        \{\O\} & \text{for any $i≥0$.}
      \end{cases}
    \]
  \end{itemize}
\end{proposition}
\begin{proof}
  See~\cite[Coro.~6.4]{silverman:elliptic}. For the characteristic $0$
  case see also next part.
\end{proof}

For curves defined over a field of positive characteristic $p$, the
case $E[p]≃ℤ/pℤ$ is called \emph{ordinary}, while the case
$E[p]≃\{\O\}$ is called \emph{supersingular}.

The free part of the group is much harder to characterize. %
We have some partial results for elliptic curves over number fields.

\begin{theorem}[Mordell-Weil]
  Let $k$ be a number field, the group $E(k)$ is finitely generated.
\end{theorem}

However the exact determination of the rank of $E(k)$ is somewhat
elusive: we have algorithms to compute the rank of most elliptic
curves over number fields; however, an exact formula for such rank is
the object of the
\href{https://en.wikipedia.org/wiki/Birch_and_Swinnerton-Dyer_conjecture}{\it
  Birch and Swinnerton-Dyer conjecture}, one of the
\href{https://en.wikipedia.org/wiki/Millennium_Prize_Problems}{\it
  Clay Millenium Prize Problems}.

\section{Maps between elliptic curves}

Finally, we focus on maps between elliptic curves. %
We are mostly interested in maps that preserve both facets of elliptic
curves: as projective varieties, and as groups. %

We first look into invertible algebraic maps, that is linear changes
of coordinates that preserve the Weierstrass form of the equation. %
Because linear maps preserve lines, it is immediate that they also
preserve the group law. %
It is easily verified that the only such maps take the form
\[(x,y) \mapsto (u^2x', u^3y')\] %
for some $u∈\bar{k}$, thus defining an \emph{isomorphism} between the
curve $y^2=x^3+au^4x+bu^6$ and the curve $(y')^2 = (x')^3 + ax' +
b$. %
Isomorphism classes are traditionally encoded by an invariant, which
origins can be tracked back to complex analysis.

\begin{proposition}[$j$-invariant]
  \label{th:j}
  Let $E:y^2=x^3+ax+b$ be an elliptic curve, and define the
  \emph{$j$-invariant} of $E$ as
  \[j(E) = 1728\frac{4a^3}{4a^3+27b^2}.\] %
  Two curves are isomorphic over the algebraic closure $\bar{k}$ if
  and only if they have the same $j$-invariant.
\end{proposition}

Note that if two curves defined over $k$ are isomorphic over
$\bar{k}$, they are so over an extension of $k$ of degree dividing
$6$. %
An isomorphism between two elliptic curves defined over $k$, that is
itself not defined over $k$ is called a \emph{twist}. %
Any curve has a \emph{quadratic twist}, unique up to isomorphism,
obtained by taking $u∉k$ such that $u^2∈k$. %
The two curves of $j$-invariant $0$ and $1728$ also have \emph{cubic},
\emph{sextic} and \emph{quartic twists}.

A surjective group morphism, not necessarily invertible, between two
elliptic curves is called an \emph{isogeny}. %
It turns out that isogenies are algebraic maps as well.

\begin{theorem}
  Let $E,E'$ be two elliptic curves, and let $\phi:E→E$ be a map between
  them. %
  The following conditions are equivalent:
  \begin{enumerate}
  \item $\phi$ is a surjective group morphism,
  \item $\phi$ is a group morphism with finite kernel,
  \item $\phi$ is a non-constant algebraic map of projective varieties
    sending the point at infinity of $E$ onto the point at infinity of
    $E'$.
  \end{enumerate}
\end{theorem}
\begin{proof}
  See~\cite[III, Th.~4.8]{silverman:elliptic}.
\end{proof}

Two curves are called \emph{isogenous} if there exists an isogeny
between them. %
We shall see in the next part that this is an equivalence relation.

Isogenies from a curve to itself are called \emph{endomorphisms}. %
The prototypical endomorphism is the multiplication-by-$m$
endomorphism defined by
\[[m]\;:\; P \mapsto [m]P.\] %
Its kernel is exactly the $m$-th torsion subgroup $E[m]$. %
For most elliptic curves, this is the end of the story: the only
endomorphisms are the scalar multiplications. %
We shall however see some non-trivial endomorphisms soon.

\section{Elliptic curves over finite fields}

From now on we let $E$ be an elliptic curve defined over a finite
field $k$ with $q$ elements. %
Obviously, the group of $k$-rational points is finite, thus the
algebraic group $E(\bar{k})$ only contains torsion elements, and we
have already characterized precisely the structure of the torsion part
of $E$.

Curves over finite fields always have a special endomorphism.

\begin{definition}[Frobenius endomorphism]
  Let $E$ be an elliptic curve defined over a field with $q$ elements,
  its \emph{Frobenius endomorphism}, denoted by $π$, is the map that
  sends
  \[(X:Y:Z) \mapsto (X^q:Y^q:Z^q).\]
\end{definition}

\begin{proposition}
  \label{th:frob}
  Let $π$ be the Frobenius endomorphism of $E$. Then:
  \begin{itemize}
  \item $\ker π = \{\O\}$;
  \item $\ker (π-1) = E(k)$.
  \end{itemize}
\end{proposition}

\begin{corollary}[Hasse's theorem]
  Let $E$ be an elliptic curve defined over a finite field $k$ with $q$
  elements, then
  \[|\#E(k) - q - 1| ≤ 2\sqrt{q}.\]
\end{corollary}
\begin{proof}
  See~\cite[V, Th.~1.1]{silverman:elliptic}.
\end{proof}

It turns out that the cardinality of $E$ over its \emph{base field}
$k$ determines its cardinality over any finite extension of it. %
This is a special case of a special case of the famous \emph{Weil's
  conjectures}, proven by Weil himself in 1949 for Abelian varieties,
and more generally by Deligne in 1973.

\begin{definition}
  Let $V$ be a projective variety defined over a finite field $\F_q$,
  its \emph{zeta function} is the power series
  \[Z(V/\F_q; T) = \exp\left(\sum_{n=1}^∞\#V(\F_{q^n})\frac{T^n}{n}\right).\]
\end{definition}

\begin{theorem}
  \label{th:weil}
  Let $E$ be an elliptic curve defined over a finite field
  $\F_q$, and let $\#E(\F_q)=q+1-a$. Then
  \[Z(E/\F_q;T) = \frac{1-aT+qT^2}{(1-T)(1-qT)}.\]
\end{theorem}
\begin{proof}
  See~\cite[V, Th.~2.4]{silverman:elliptic}.
\end{proof}

We conclude with a theorem that links the isogenies between two
elliptic curves with their Frobenius endomorphisms.

\begin{theorem}[Sato-Tate]
  \label{th:sato-tate}
  Two elliptic curves $E,E'$ defined over a finite field $k$ are
  isogenous over $k$ if and only if $\#E(k)=\#E'(k)$.
\end{theorem}

\section{Application: Diffie-Hellman key exhange}
\label{sec:appl-diff-hellm}

Elliptic curves are largely present in modern technology thanks to
their applications in cryptography. %
The simplest of these application is the \emph{Diffie-Hellman key
  exchange}, a cryptographic protocol by which two parties
communicating over a public channel can agree on a common secret
string unknown to any other party listening on the same channel.

The original protocol was invented in the 1970s by Whitfield Diffie
and Martin Hellman~\cite{dh}, and constitutes the first practical
example of \emph{public key cryptography}. %
The two communicating parties are customarily called \emph{Alice} and
\emph{Bob}, and the listening third party is represented by the
character \emph{Eve} (for \emph{eavesdropper}). %
To set up the protocol, Alice and Bob agree on a set of public
parameters:
\begin{itemize}
\item A \emph{large enough} prime number $p$, such that $p-1$ has a
  \emph{large enough} prime factor;
\item A multiplicative generator $g∈ℤ/pℤ$.
\end{itemize}

Then, Alice and Bob perform the following steps:
\begin{enumerate}
\item Each chooses a \emph{secret} integer in the interval $]0,p-1[$;
  call $a$ \emph{Alice's secret} and $b$ \emph{Bob's secret}.
\item They respectively compute $A=g^a$ and $B=g^b$.
\item They exchange $A$ and $B$ over the public channel.
\item They respectively compute the \emph{shared secret}
  $B^a=A^b=g^{ab}$.
\end{enumerate}

The protocol can be easily generalized by replacing the multiplicative
group $(ℤ/pℤ)^{×}$ with any other cyclic group $G=〈g〉$. %
From Eve's point of view, she is given the knowledge of the group $G$,
the generator $g$, and Alice's and Bob's public data $A,B∈G$; her goal
is to recover the shared secret $g^{ab}$. %
This is mathematically possible, but not necessarily \emph{easy} from
a computational point of view.

\begin{definition}[Discrete logarithm]
  Let $G$ be a cyclic group generated by an element $g$. For any
  element $A∈G$, we define the \emph{discrete logarithm of $A$ in base
    $g$}, denoted $\log_g(A)$, as the unique integer in the interval
  $[0,\#G[$ such that
  \[g^{\log_g(A)} = A.\]
\end{definition}

It is evident that if Eve can compute discrete logarithms in $G$
efficiently, then she can also efficiently compute the shared secret;
the converse is not known to be true in general, but it is widely
believed to be. %
Thus, the strength of the Diffie-Hellman protocol is entirely
dependent on the \emph{hardness} of the \emph{discrete logarithm
  problem} in the group $G$.

We know algorithms to compute discrete logarithms in a \emph{generic}
group $G$ that require $O(\sqrt{q})$ computational steps
(see~\cite{joux2009algorithmic}), where $q$ is the largest prime
divisor of $\#G$; we also know that these algorithms are \emph{optimal
  for abstract cyclic groups}. %
For this reason, $G$ is usually chosen so that the largest prime
divisor $q$ has size at least $\log_2 q ≈ 256$. %
However, the proof of optimally does not exclude the existence of
better algorithms for \emph{specific} groups $G$. %
And indeed, algorithms of complexity better than $O(\sqrt{\#G})$ are
known for the case $G=(ℤ/pℤ)^{×}$~\cite{joux2009algorithmic}, thus
requiring parameters of considerably larger size to guarantee
cryptographic strength.

On the contrary, no algorithms better than the generic ones are known
when $G$ is a subgroup of $E(k)$, where $E$ is an elliptic curve
defined over a finite field $k$. %
This has led Miller~\cite{miller86} and Koblitz~\cite{koblitz87} to
suggest, in the 1980s, to replace $(ℤ/pℤ)^{×}$ in the Diffie-Helman
protocol by the group of rational points of an elliptic curve of
(almost) prime order over a finite field. %
The resulting protocol is summarized in Figure~\ref{fig:dh}.

\begin{figure}
  \centering
  \begin{tabular}{l *{2}{p{25ex}<{\centering}}}
    \hline
    Public parameters & \multicolumn{2}{l}{Finite field $\F_p$, with $\log_2p≈256$,}\\
                      & \multicolumn{2}{l}{Elliptic curve $E/\F_p$, such that $\#E(\F_p)$ is prime,}\\
                      & \multicolumn{2}{l}{A generator $P$ of $E(\F_p)$.}\\
    \hline
                      & {\bf Alice} & {\bf Bob}\\
    \hline
    Pick random secret & $0<a<\#E(\F_p)$ & $0<b<\#E(\F_p)$\\
    Compute public data & $A = [a]P$ & $B = [b]P$\\
    Exchange data &  \hfill $A \longrightarrow$ & $\longleftarrow B$ \hfill\strut \\
    Compute shared secret & $S = [a]B$ & $S = [b]A$
  \end{tabular}
  
  \caption{The Diffie-Hellman protocol over elliptic curves}
  \label{fig:dh}
\end{figure}

\section{Application: Elliptic curve factoring method}

A second popular use of elliptic curves in technology is for factoring
large integers, a problem that also occurs frequently in cryptography.

The earliest method for factoring integers was already known to the
ancient Greeks: the \emph{sieve of Eratosthenes} finds all primes up
to a given bound by crossing composite numbers out in a table. %
Applying the Eratosthenes' sieve up to $\sqrt{N}$ finds all prime
factors of a composite number $N$. %
Examples of modern algorithms used for factoring are Pollard's
\emph{Rho algorithm} and Coppersmith's \emph{Number Field Sieve
  (NFS)}.

In the 1980s H. Lenstra~\cite{lenstra87} introduced an algorithm for
factoring that has become known as the \emph{Elliptic Curve Method
  (ECM)}. %
Its complexity is between Pollard's and Coppersmith's algorithms in
terms of number of operations; at the same time it only requires a
constant amount of memory, and is very easy to parallelize. %
For these reasons, ECM is typically used to factor integers having
medium sized prime factors.

From now on we suppose that $N=pq$ is an integer which factorization
we wish to compute, where $p$ and $q$ are distinct primes. %
Without loss of generality, we can suppose that $p<q$.

Lenstra's idea has its roots in an earlier method for factoring
special integers, also due to Pollard. %
Pollard's \emph{$(p-1)$ factoring method} is especially suited for
integers $N=pq$ such that $p-1$ only has \emph{small} prime factors. %
It is based on the isomorphism
\begin{align*}
  \rho : ℤ/Nℤ &\to ℤ/pℤ × ℤ/qℤ,\\
  x &\mapsto (x \bmod p, x \bmod q)
\end{align*}
given by the Chinese remainder theorem. %
The algorithm is detailed in Figure~\ref{fig:p-1}. %
It works by guessing a multiple $e$ of $p-1$, then taking a random
element $x∈(ℤ/Nℤ)^{×}$, to deduce a random element $y$ in
$〈1〉⊕(ℤ/qℤ)^{×}$. If the guessed exponent $e$ was correct, and if
$y≠1$, the gcd of $y-1$ with $N$ yields a non-trivial factor. %

\begin{figure}
  \begin{subfigure}{0.45\textwidth}
    \begin{algorithmic}[1]
      \REQUIRE An integer $N=pq$,\\
      a bound $B$ on the largest prime factor of $p-1$;
      \ENSURE $(p,q)$ or FAIL.
      \STATE Set $e = \prod_{r \text{ prime } < B} r^{\lfloor\log_r\sqrt{N}\rfloor}$;
      \STATE Pick a random $1 < x < N$;
      \STATE Compute $y = x^e \mod N$;
      \STATE Compute $q' = \gcd(y-1, N)$;
      \IF {$q' ≠ 1,N$}
      \RETURN $N/q', q'$;
      \ELSE
      \RETURN FAIL.
      \ENDIF
    \end{algorithmic}
    
    \caption{Pollard's $(p-1)$ algorithm}
    \label{fig:p-1}
  \end{subfigure}
  \hfill
  \begin{subfigure}{0.45\textwidth}
    \begin{algorithmic}[1]
      \REQUIRE An integer $N=pq$, a bound $B$;
      \ENSURE $(p,q)$ or FAIL.
      \STATE Pick random integers $a,X,Y$ in $[0,N[$;
      \STATE Compute $b = Y^2 - X^3 - aX \mod N$;
      \STATE Define the elliptic curve $E \;:\; y^2 = x^3 - ax - b$.
      \STATE Define the point $P=(X:Y:1) ∈ E(ℤ/Nℤ)$.
      \STATE Set $e = \prod_{r \text{ prime } < B} r^{\lfloor\log_r\sqrt{N}\rfloor}$;
      \STATE Compute $Q = [e]P = (X':Y':Z')$;
      \STATE Compute $q' = \gcd(Z', N)$;
      \IF {$q' ≠ 1,N$}
      \RETURN $N/q', q'$;
      \ELSE
      \RETURN FAIL.
      \ENDIF
    \end{algorithmic}
    
    \caption{Lenstra's ECM algorithm}
    \label{fig:ecm}
  \end{subfigure}
  \caption{The $(p-1)$ and ECM factorization algorithms}
\end{figure}

The $p-1$ method is very effective when the bound $B$ is small, but
its complexity grows exponentially with $B$. %
For this reason it is only usable when $p-1$ has small prime factors,
a constraint that is very unlikely to be satisfied by random primes.

Lenstra's ECM algorithm is a straightforward generalization of the
$p-1$ method, where the multiplicative groups $(ℤ/pℤ)^{×}$ and
$(ℤ/qℤ)^{×}$ are replaced by the groups of points $E(\F_p)$ and
$E(\F_q)$ of an elliptic curve defined over $ℚ$. %
Now, the requirement is that $\#E(\F_p)$ only has small prime
factors. %
This condition is also extremely rare, but now we have the freedom to
try the method many times by changing the elliptic curve. %

The algorithm is summarized in Figure~\ref{fig:ecm}. %
It features two remarkable subtleties. %
First, it would feel natural to pick a random elliptic curve
$E:y^2=x^3+ax+b$ by picking random $a$ and $b$, however taking a point
on such curve would then require computing a square root modulo $N$, a
problem that is known to be has hard as factoring $N$. %
For this reason, the algorithm starts by taking a random point, and
then deduces the equation of $E$ from it. %
Secondly, all computations on coordinates happen in the projective
plane over $ℤ/Nℤ$; however, properly speaking, projective space cannot
be defined over non-integral rings. %
Implicitly, $E(ℤ/Nℤ)$ is defined as the product group
$E(\F_p)⊕E(F_q)$, and any attempt at inverting a non-invertible in
$ℤ/Nℤ$ will result in a factorization of $N$.

\section*{Exercices}

\begin{exercice}
  Prove Proposition~\ref{th:j}.
\end{exercice}

\begin{exercice}
  Determine all the possible automorphisms of elliptic curves.
\end{exercice}

\begin{exercice}
  Prove Proposition~\ref{th:frob}.
\end{exercice}

\begin{exercice}
  Using Proposition~\ref{th:weil}, devise an algorithm to effectively
  compute $\#E(\F_{q^n})$ given $\#E(\F_q)$.
\end{exercice}

\begin{exercice}
  Implement the ECDH key exchange in the language of your choice.
\end{exercice}

\begin{exercice}[Pohlig-Hellman algorithm]
  Let $G$ be a cyclic group of order $N=pq$, generated by an element
  $g$. %
  Show how to solve discrete logarithms in $G$ by computing two
  separate discrete logarithms in the subgroups $〈g^p〉$ and $〈g^q〉$.
\end{exercice}

\begin{exercice}
  Implement the ECM factorization method in the language of your
  choice.
\end{exercice}


\clearpage
\part{Isogenies and applications}

\section{Elliptic curves over $ℂ$}

\begin{definition}[Complex lattice]
  A complex lattice $Λ$ is a discrete subgroup of $ℂ$ that contains an
  $ℝ$-basis.
\end{definition}

Explicitly, a complex lattice is generated by a \emph{basis}
$(ω_1,ω_2)$, such that $ω_1≠λω_2$ for any $λ∈ℝ$, as
\[Λ = ω_1ℤ + ω_2ℤ.\] %
Up to exchanging $ω_1$ and $ω_2$, we can assume that $\im(ω_1/ω_2)>0$;
we then say that the basis has \emph{positive orientation}. %
A positively oriented basis is obviously not unique, though.

\begin{proposition}
  \label{th:basis-change}
  Let $Λ$ be a complex lattice, and let $(ω_1,ω_2)$ be a positively
  oriented basis, then any other positively oriented basis
  $(ω_1',ω_2')$ is of the form
  \begin{align*}
    ω_1' &= aω_1 + bω_2,\\
    ω_1' &= cω_1 + dω_2,
  \end{align*}
  for some matrix
  $\left(\begin{smallmatrix}a&b\\c&d\end{smallmatrix}\right)∈\SL_2(ℤ)$.
\end{proposition}
\begin{proof}
  See~\cite[I, Lem.~2.4]{silverman:advanced}.
\end{proof}

\begin{definition}[Complex torus]
  Let $Λ$ be a complex lattice, the quotient $ℂ/Λ$ is called a
  \emph{complex torus}.
\end{definition}

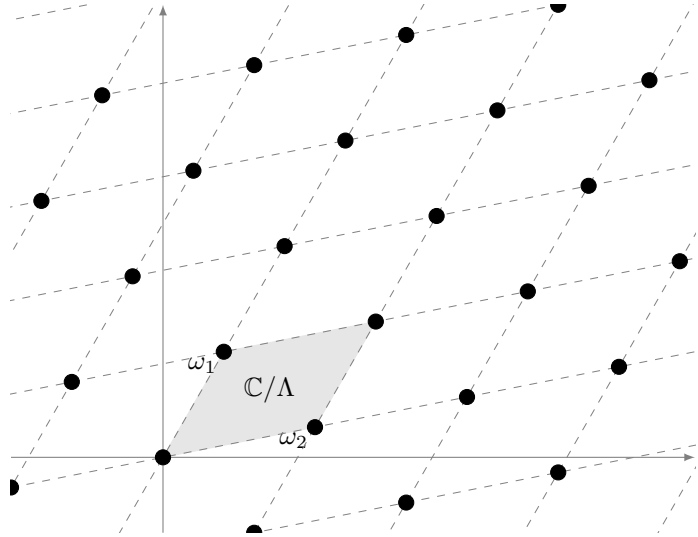
\begin{figure}
  \centering
  \begin{tikzpicture}[scale=2]
    \axes{-1}{3.5}{-0.5}{3}

    \begin{scope}[/lattice={1}{0.2}{0.4}{0.7}]
      \draw[fill,black!10] (0,0) -- (1,0) -- (1,1) -- (0,1) -- (0,0);
      \node at (0.5,0.5) {$ℂ/Λ$};
      \node at (0.9,-0.1) {$ω_2$};
      \node at (-0.1,0.9) {$ω_1$};

      \lattice{-3}{4}
    \end{scope}  
  \end{tikzpicture}

  \caption{A complex lattice (black dots) and its associated complex
    torus (grayed \emph{fundamental domain}).}
  \label{fig:lattice}
\end{figure}

\begin{figure}
  \centering
  \begin{tikzpicture}[scale=1.8]
    \axes{-0.5}{3.5}{-0.5}{3}

    \begin{scope}[/lattice={1}{0.2}{0.4}{0.7}]
      \lattice{-3}{4}

      \node[red] at (0.7,0.65) {$a$}; 
      \node[draw,circle,inner sep=1pt,fill,red] at (0.8,0.5) {};
      \node[red] at (0.2,0.9) {$b$}; 
      \node[draw,circle,inner sep=1pt,fill,red] at (0.3,0.7) {};
      
      \node[draw,circle,inner sep=1pt,fill,red] at (1.1,1.2) {};

      \draw[red,thin,dotted] (0,0) -- (0.8,0.5) -- (1.1,1.2)
      (0,0) -- (0.3,0.7) -- (1.1,1.2);          

      \node[red] at (0.2,0.3) {$a+b$}; 
      \node[draw,circle,inner sep=1pt,fill,red] at (0.1,0.2) {};
    \end{scope}  
  \end{tikzpicture}
  \hfill
  \begin{tikzpicture}[scale=1.8]
    \axes{-0.5}{3.5}{-0.5}{3}

    \begin{scope}[/lattice={1}{0.2}{0.4}{0.7}]
      \lattice{-3}{4}
      
      \node[red,yshift=0.2cm] at (0.8,0.6) {$a$}; 
      \draw[red] (0.8,0.6) node[fill,circle,inner sep=1pt] {};

      \draw[red,dotted] (0,0) -- (1.6,1.2) node[fill,circle,inner sep=1pt] {} 
      -- (2.4,1.8) node[fill,circle,inner sep=1pt] {};

      \node[red,yshift=0.3cm] at (0.4,0.8) {$[3]a$}; 
      \draw[red] (0.4,0.8) node[fill,circle,inner sep=1pt] {};
    \end{scope}
  \end{tikzpicture}
  \caption{Addition (left) and scalar multiplication (right) of points
    in a complex torus $ℂ/Λ$.}
  \label{fig:lattice-arith}
\end{figure}
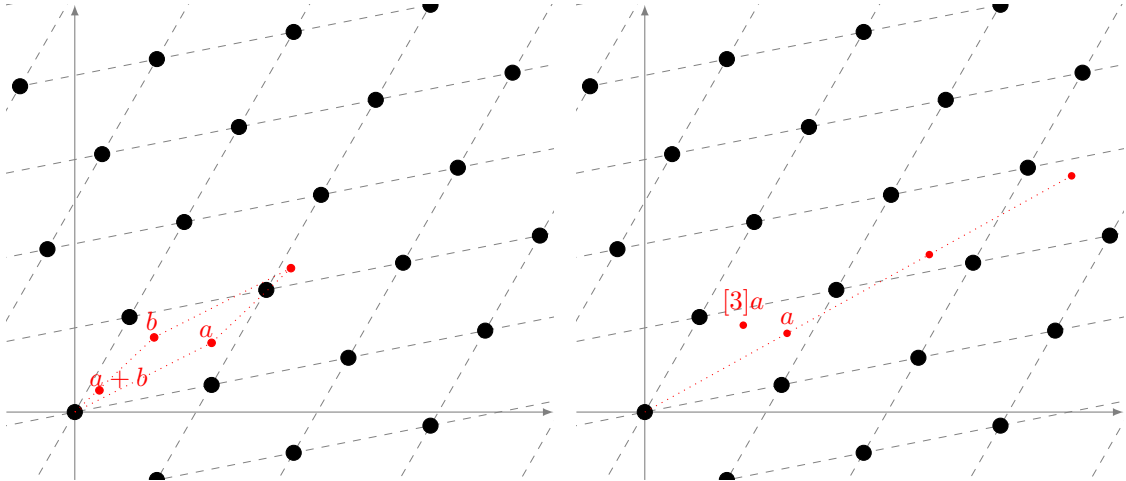

A convex set of class representatives of $ℂ/Λ$ is called a
\emph{fundamental parallelogram}. %
Figure~\ref{fig:lattice} shows a complex lattice generated by a
(positively oriented) basis $(ω_1,ω_2)$, together with a fundamental
parallelogram for $ℂ/(ω_1,ω_2)$. %
The additive group structure of $ℂ$ carries over to $ℂ/Λ$, and can be
graphically represented as operations on points inside a fundamental
parallelogram. %
This is illustrated in Figure~\ref{fig:lattice-arith}.

\begin{definition}[Homothetic lattices]
  Two complex lattices $Λ$ and $Λ'$ are said to be \emph{homothetic}
  if there is a complex number $α∈ℂ^{×}$ such that $Λ=αΛ'$.
\end{definition}

Geometrically, applying a homothety to a lattice corresponds to zooms
and rotations around the origin. %
We are only interested in complex tori up to homothety; to classify
them, we introduce the \emph{Eisenstein series of weight $2k$},
defined as
\[G_{2k}(Λ) = \sum_{ω∈Λ\setminus\{0\}}ω^{-2k}.\]
It is customary to set
\[g_2(Λ) = 60G_4(Λ), \quad g_3(Λ) = 140G_6(Λ);\] %
when $Λ$ is clear from the context, we simply write $g_2$ and $g_3$.

\begin{theorem}[Modular $j$-invariant]
  The \emph{modular $j$-invariant} is the function on complex lattices
  defined by
  \[j(Λ) = 1728 \frac{g_2(Λ)^3}{g_2(Λ)^3 - 27g_3(Λ)^2}.\] %
  Two lattices are homothetic if and only if they have the same
  modular $j$-invariant.
\end{theorem}
\begin{proof}
  See~\cite[I, Th.~4.1]{silverman:advanced}.
\end{proof}

It is no chance that the invariants classifying elliptic curves and
complex tori look very similar. %
Indeed, we can prove that the two are in one-to-one correspondence.

\begin{definition}[Weierstrass $℘$ function]
  Let $Λ$ be a complex lattice, the \emph{Weierstrass $℘$ function}
  associated to $Λ$ is the series
  \[℘(z;Λ) = \frac{1}{z^2} + \sum_{ω∈Λ\setminus\{0\}} \left(\frac{1}{(z-ω)^2} - \frac{1}{ω^2}\right).\]
\end{definition}

\begin{theorem}
  The Weierestrass function $℘(z;Λ)$ has the following properties:
  \begin{enumerate}
  \item It is an \emph{elliptic function} for $Λ$, i.e.
    $℘(z) = ℘(z+ω)$ for all $z∈ℂ$ and $ω∈Λ$.
  \item Its Laurent series around $z=0$ is
    \[℘(z) = \frac{1}{z^2} + \sum_{k=1}^∞(2k+1)G_{2k+2}z^{2k}.\]
  \item It satisfies the differential equation
    \[℘'(z)^2 = 4℘(z)^3 - g_2℘(z) - g_3\]
    for all $z∉Λ$.
  \item The curve
    \[E\;:\;y^2=4x^3 - g_2x - g_3\]
    is an elliptic curve over $ℂ$. The map
    \begin{align*}
      ℂ/Λ &\to E(ℂ),\\
      0 &\mapsto (0:1:0),\\
      z &\mapsto (℘(z):℘'(z):1)
    \end{align*}
    is an isomorphism of Riemann surfaces and a group morphism.
  \end{enumerate}
\end{theorem}
\begin{proof}
  See~\cite[VI, Th.~3.1, Th.~3.5, Prop.~3.6]{silverman:elliptic}.
\end{proof}

By comparing the two definitions for the $j$-invariants, we see that
$j(Λ)=j(E)$. %
So, for any homotety class of complex tori, we have a corresponding
isomorphism class of elliptic curves. %
The converse is also true.

\begin{theorem}[Uniformization theorem]
  Let $a,b∈ℂ$ be such that $4a^3+27b^2≠0$, then there is a unique
  complex lattice $Λ$ such that $g_2(Λ) = -4a$ and $g_3(Λ) = -4b$.
\end{theorem}
\begin{proof}
  See~\cite[I, Coro.~4.3]{silverman:advanced}.
\end{proof}

\begin{figure}

  \begin{subfigure}{.45\textwidth}
    \centering
    
    \begin{tikzpicture}[scale=1.2]
      \axes{-0.3}{4.5}{-0.5}{4};

      \begin{scope}[/lattice={3}{0.6}{1.2}{2.1}]
        \lattice{-1}{2}

        \foreach \i in {0,...,2} {
          \foreach \j in {0,...,2} {
            \draw[red] (\i/3,\j/3) node[fill,circle,inner sep=1pt] {};
          }
        }
        \draw[red] (0,0) -- (1/3,0) node[yshift=0.2cm] {$a$};
        \draw[red] (0,0) -- (0,1/3) node[yshift=0.2cm] {$b$};

        \draw[blue] (0.8,0.5) node[fill,circle,inner sep=1pt] {}
        node[yshift=0.2cm] {\scriptsize $z$}
        (2/15,1/6) node[fill,circle,inner sep=1pt] {}
        node[yshift=0.2cm] {\scriptsize $3z$};
      \end{scope}
    \end{tikzpicture}  
    \caption{$3$-torsion group on a complex torus (red
      points), with two generators $a$ and $b$, and action of the
      multiplication-by-$3$ map (blue dots).}
    \label{fig:torsion}
  \end{subfigure}
  \hfill
  \begin{subfigure}{.45\textwidth}
    \centering
    \begin{tikzpicture}[scale=1.2]
      \axes{-0.3}{4.5}{-0.5}{4};
      
      \begin{scope}[/lattice={3}{0.6}{1.2}{2.1}]
        \lattice{-1}{2}

        \draw[red] (0,0) -- (1/3,0) node[yshift=0.3cm] {$a$};
        \draw[green] (0,0) -- (0,1/3) node[fill,circle,inner sep=1pt] {}
        node[yshift=0.3cm] {$b$};

        \draw[blue] (0.8,0.5) node[fill,circle,inner sep=1pt] {}
        node[yshift=0.3cm] {$z$};
      \end{scope}
      
      \begin{scope}[/lattice={1}{0.2}{1.2}{2.1}]
        \begin{scope}[opacity=0.5,red]
          \lattice[1pt]{-3}{5}
        \end{scope}

        \draw[blue] (0.4,0.5) node[fill,circle,inner sep=1pt] {}
        node[yshift=0.3cm] {$ϕ(z)$};
      \end{scope}
    \end{tikzpicture}
    
    \caption{Isogeny from $ℂ/Λ$ (black dots) to $ℂ/Λ'$ (red dots)
      defined by $ϕ(z)=z \bmod Λ'$. The kernel of $ϕ$ is contained
      in $(ℂ/Λ)[3]$ and is generated by $a$. The kernel of the dual
      isogeny $\hat{ϕ}$ is generated by the vector $b$ in $Λ'$.}
    \label{fig:isogeny}
  \end{subfigure}
  
  \caption{Maps between complex tori.}
\end{figure}
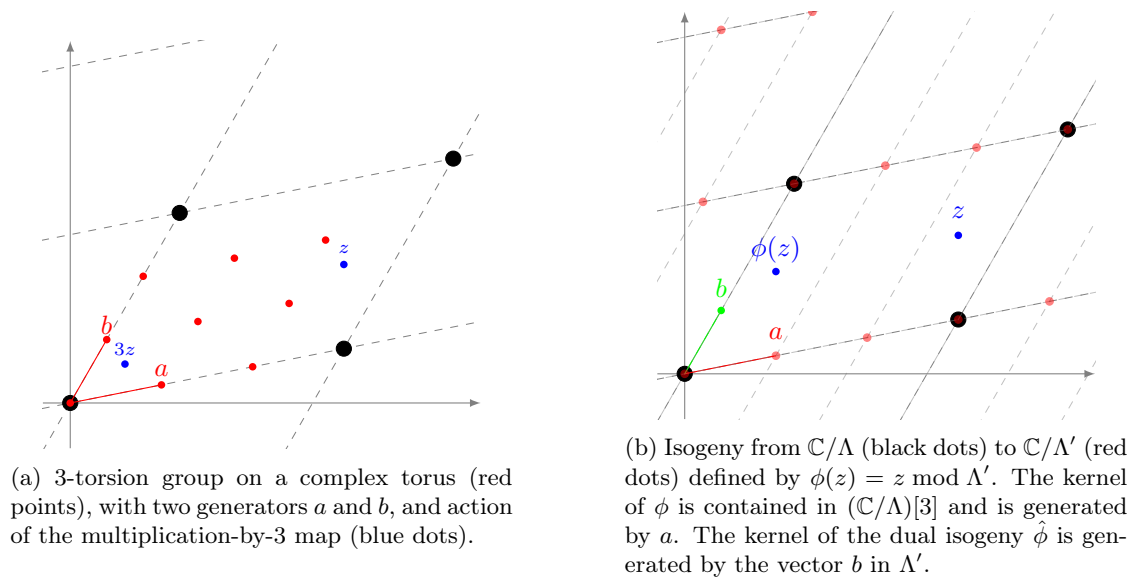

Using the correspondence between elliptic curves and complex tori, we
now have a new perspective on their group structure. %
Looking at complex tori, it becomes immediately evident why the
torsion part has rank $2$, i.e. why $E[m]≃(ℤ/mℤ)^2$. %
This is illustrated in Figure~\ref{fig:torsion}; in the picture wee
see two lattices $Λ$ and $Λ'$, generated respectively by the black and
the red dots. %
The multiplication-by-$m$ map corresponds then to
\begin{align*}
  [m] : ℂ/Λ &\to ℂ/Λ',\\
  z &\mapsto z \bmod Λ';
\end{align*}
and we verify that it is an endomorphism because $Λ$ and $Λ'$ are
homothetic.

Within this new perspective, isogenies are a mild generalization of
scalar multiplications. %
Whenever two lattices $Λ,Λ'$ verify $αΛ⊂Λ'$, there is a well defined
map
\begin{align*}
   ϕ_α : ℂ/Λ &\to ℂ/Λ',\\
  z &\mapsto αz \bmod Λ'
\end{align*}
that is holomorphic and also a group morphism. %
One example of such maps is given in Figure~\ref{fig:torsion}: there,
$α=1$ and the red lattice strictly contains the black one; the map is
simply defined as reduction modulo $Λ'$. %
It turns out that these maps are exactly the isogenies of the
corresponding elliptic curves.

\begin{theorem}
  Let $E,E'$ be elliptic curves over $ℂ$, with corresponding lattices
  $Λ,Λ'$. %
  There is a bijection between the group of isogenies from $E$ to $E'$
  and the group of maps $ϕ_α$ for all $α$ such that $Λ⊂αΛ'$.
\end{theorem}
\begin{proof}
  See~\cite[VI, Th.~4.1]{silverman:elliptic}.
\end{proof}

Looking again at Figure~\ref{fig:torsion}, we see that there is a
second isogeny $\hat{ϕ}$ from $Λ'$ to $Λ/3$, which kernel is generated
by $b∈Λ'$. %
The composition $\hat{ϕ}∘ϕ$ is an endomorphism of $ℂ/Λ$, up to the
homothety sending $Λ/3$ to $Λ$, and we verify that it corresponds to
the multiplication-by-$3$ map. %
In this example, the kernels of both $ϕ$ and $\hat{ϕ}$ contain $3$
elements, and we say that $ϕ$ and $\hat{ϕ}$ have \emph{degree} $3$. %
Although not immediately evident from the picture, this same
construction can be applied to any isogeny. %
The isogeny $\hat{ϕ}$ is called the \emph{dual} of $ϕ$. %
Dual isogenies exist not only in characteristic $0$, but for any base
field. %

We finish this section by summarizing the most important algebraic
properties of isogenies; we start with a technical definition.

\begin{definition}[Degree, separability]
  Let $ϕ:E\to E'$ be an isogeny defined over a field $k$, and let
  $k(E),k(E')$ be the function fields of $E,E'$. %
  By composing $\phi$ with the functions of $k(E')$, we obtain a
  subfield of $k(E)$ that we denote by $ϕ^\ast(k(E'))$.

  \begin{enumerate}
  \item The \emph{degree} of $ϕ$ is defined as
    $\deg ϕ = [k(E):ϕ^\ast(k(E'))]$; it is always finite.
  \item $ϕ$ is said to be \emph{separable}, \emph{inseparable}, or
    \emph{purely inseparable} if the extension of function fields is.
  \item If $ϕ$ is separable, then $\deg ϕ = \#\ker ϕ$.
  \item If $ϕ$ is purely inseparable, then $\deg ϕ$ is a power of the
    characteristic of $k$.
  \item Any isogeny can be decomposed as a product of a separable and
    a purely inseparable isogeny.
  \end{enumerate}
\end{definition}
\begin{proof}
  See~\cite[II, Th.~2.4]{silverman:elliptic}.
\end{proof}

In practice, most of the time we will be considering separable
isogenies, and we can take $\deg ϕ = \#\ker ϕ$ as the definition of
the degree. %
Notice that in this case $\deg ϕ$ is the size of any fiber of $ϕ$. %
Separable isogenies are completely determined by their kernel, as the
following proposition shows.

\begin{proposition}
  Let $E$ be an elliptic curve, and let $G$ be a finite subgroup of
  $E$. %
  There are a unique elliptic curve $E'$, and a unique separable
  isogeny $ϕ$, such that $\ker ϕ=G$ and $ϕ:E\to E'$. %
\end{proposition}
\begin{proof}
  See~\cite[Prop.~III, 4.12]{silverman:elliptic}.
\end{proof}

The proposition justifies introducing the notation $E/G$ for the image
curve $E'$. %
We conclude with a fundamental theorem on isogenies.

\begin{theorem}[Dual isogeny]
  Let $ϕ:E\to E'$ be an isogeny of degree $m$. %
  There is a unique isogeny $\hat{ϕ}:E'\to E$ such that
  \[\hat{ϕ}∘ϕ = [m]_E, \quad ϕ∘\hat{ϕ} = [m]_{E'}.\] %
  $\hat{ϕ}$ is called the \emph{dual isogeny of $ϕ$}; it has the
  following properties:
  
  \begin{enumerate}
  \item $\hat{ϕ}$ is defined over $k$ if and only if $ϕ$ is;
  \item $\widehat{ψ∘ϕ} = \hat{ϕ}∘\hat{ψ}$ for any isogeny $ψ:E'\to E''$;
  \item $\widehat{ψ+ϕ} = \hat{ψ} + \hat{ϕ}$ for any isogeny $ψ:E\to E'$;
  \item $\deg ϕ = \deg\hat{ϕ}$;
  \item $\hat{\hat{ϕ}} = ϕ$.
  \end{enumerate}
\end{theorem}

\section{The endomorphism ring}

We have already defined an endomorphism as an isogeny from a curve to
itself. %
If we add the multiplication-by-$0$ to it, the set of all
endomorphisms of $E$ form a ring under the operations of addition and
composition, denoted by $\End(E)$. %

We have already seen that the multiplication-by-$m$ is a different
endomorphism for any integer $m$, thus $ℤ⊂\End(E)$. %
For the case of finite fields, we have also learned about the
Frobenius endomorphism $π$; so certainly $ℤ[π]⊂\End(E)$ in this
case. %
We shall now give a complete characterization of the endomorphism ring
for any field.

\begin{definition}[Order]
  Let $K$ be a finitely generated $ℚ$-algebra. %
  An \emph{order} $\O⊂K$ is a subring of $K$ that is a finitely
  generated $ℤ$-module of maximal dimension.
\end{definition}

The prototypical example of order is the ring of integers $\O_K$ of a
number field $K$, i.e., the ring of all elements of $K$ such that
their monic minimal polynomial has coefficients in $ℤ$. %
It turns out that $\O_K$ is the \emph{maximal order} of $K$, i.e., it
contains any other order of $K$.

\begin{definition}[Quaternion algebra]
  A \emph{quaternion algebra} is an algebra of the form
  \[K = ℚ + αℚ + βℚ + αβℚ,\]
  where the generators satisfy the relations
  \[α^2,β^2∈ℚ, \quad α^2<0, \quad β^2 < 0, \quad βα=-αβ.\]
\end{definition}

\begin{theorem}[Deuring]
  Let $E$ be an elliptic curve defined over a field $k$ of
  characteristic $p$. %
  The ring $\End(E)$ is isomorphic to one of the following:
  \begin{itemize}
  \item $ℤ$, only if $p=0$;
  \item An order $\O$ in a quadratic imaginary field (a number field
    of the form $ℚ[\sqrt{-D}]$ for some $D>0$); in this case we say
    that $E$ has \emph{complex multiplication} by $\O$;
  \item Only if $p>0$, a maximal order in the quaternion algebra
    ramified at $p$ and $∞$; in this case we say that $E$ is
    \emph{supersingular}.
  \end{itemize}
\end{theorem}
\begin{proof}
  See~\cite[III, Coro.~9.4]{silverman:elliptic}
  and~\cite{belding08-thesis}.
\end{proof}

In positive characteristic, a curve that is not supersingular is
called \emph{ordinary}; it necessarily has complex multiplication. %
We focus again on the finite field case; we have already seen that
$Ζ[π]⊂\End(E)$. %
Now, Hasse's theorem can be made more precise as follows.

\begin{theorem}
  Let $E$ be an elliptic curve defined over a finite field. %
  Its Frobenius endomorphism $π$ satisfies a quadratic equation
  \[π^2 - tπ + q = 0,\]
  for some $|t|≤2\sqrt{q}$.
\end{theorem}
\begin{proof}
  See~\cite[V, Th.~2.3.1]{silverman:elliptic}.
\end{proof}

The coefficient $t$ in the equation is called the \emph{trace} of
$π$. %
By replacing $π=1$ in the equation, we immediately obtain the
cardinality of $E$ as $\#E = q+1-t$. %
Now, if we let $D_π=t^2-4q<0$, we verify that $π∈ℚ[\sqrt{D_π}]$; so,
at least in the ordinary case, we can affirm that
\[ℤ[π] ⊂ \End(E) ⊂ \O_K,\] %
where $K=ℚ[\sqrt{D_π}]$ is called the \emph{endomorphism algebra} of
$E$. %
The structure of the orders of $K$ is very simple in this case.

\begin{proposition}
  Let $K$ be a quadratic number field, and let $\O_K$ be its ring of
  integers. %
  Any order $\O⊂K$ can be written as $\O=ℤ+f\O_K$ for an integer $f$,
  called the \emph{conductor} of $\O$. %
  If $d_K$ is the \emph{discriminant} of $K$, the discriminant of $\O$
  is $f^2d_K$.

  If $\O,\O'$ are two orders of discriminants $f,f'$, then $\O⊂\O'$ if
  and only if $f'|f$.
\end{proposition}

In our case, we can write $D_π=f^2d_K$, with $d_K$
squarefree. 
Then, any order $ℤ[π] ⊂ \O ⊂ \O_K$ has conductor dividing $f$.

\section{Application: point counting}
\label{sec:appl-point-count}

Before going more in depth into the study of the endomorphism ring,
let us pause for a while on a simpler problem. %
Hasse's theorem relates the cardinality of a curve defined over a
finite field with the trace of its Frobenius endomorphism. %
However, it does not give us an algorithm to compute either.

The first efficient algorithm to compute the trace of $π$ was proposed
by Schoof in the 1980s~\cite{schoof85}. %
The idea is very simple: compute the value of $t_π\bmod ℓ$ for many
small primes $ℓ$, and then reconstruct the trace using the Chinese
remainder theorem. %
To compute $t_π\bmod ℓ$, Schoof's algorithm formally constructs the
group $E[ℓ]$, takes a generic point $P∈E[\ell]$, and then runs a
search for the integer $t$ such that
\[π([t]P) = [q]P + π^2(P).\] %
The formal computation must be carried out by computing modulo a
polynomial that vanishes on the whole $E[\ell]$; the smallest such
polynomial is provided by the \emph{division polynomial} $ψ_ℓ$.

\begin{definition}[Division polynomial]
  Let $E:y^2=x^3+ax+b$ be an elliptic curve, the \emph{division
    polynomials} $ψ_m$ are defined by the initial values
  \begin{align*}
    ψ_1 &= 1,\\
    ψ_2 &= 2y^2,\\
    ψ_3 &= 3x^4 + 6ax^2 + 12bx - a^2,\\
    ψ_4 &= (2x^6 + 10ax^4 + 40bx^3 - 10a^2x^2 - 8abx - 2a^3 - 16b^2)2y^2,
  \end{align*}
  and by the recurrence
  \begin{align*}
    ψ_{2m+1}  &= ψ_{m+2}ψ_m^3 - ψ_{m-1}ψ_{m+1}^3 &\text{for $m≥2$,}\\
    ψ_2ψ_{2m} &= (ψ_{m+2}ψ_{m-1}^2 - ψ_{m-2}ψ_{m+1}^2)ψ_m &\text{for $m≥3$.}
  \end{align*}

  The $m$-th division polynomial $ψ_m$ vanishes on $E[m]$;
  the multiplication-by-$m$ map can be written as
  \[[m]P = \left(\frac{ϕ_m(P)}{ψ_m(P)^2}, \frac{ω_m(P)}{ψ_m(P)^3}\right)\]
  for any point $P≠\O$, where $ϕ_m$ and $ω_m$ are defined as
  \begin{align*}
    ϕ_m &= xψ_m^2 - ψ_{m+1}ψ_{m-1},\\
    ω_m &= ψ_{m-1}^2ψ_{m+2} + ψ_{m-2}ψ_{m+1}^2.
  \end{align*}
\end{definition}

Schoof's algorithm runs in time polynomial in $\log\#E(k)$, however it
is quite slow in practice. %
Among the major advances that have enabled the use of elliptic curves
in cryptography are the optimizations of Schoof's algorithm due to
Atkin and Elkies~\cite{atkin88,atkin91,elkies92,schoof95,elkies98}. %
Both improvements use a better understanding of the action of $π$ on
$E[ℓ]$. %
Assume that $ℓ$ is different from the characteristic, we have already
seen that $E[ℓ]$ is a group of rank two. %
Hence, $π$ acts on $E[ℓ]$ like a matrix $M$ in $\GL_2(ℤ/ℓℤ)$, and its
characteristic polynomial is exactly
\[χ(X) = X^2 - t_πX + q \mod \ell.\] %
Now we have three possibilities:
\begin{itemize}
\item $χ$ splits modulo $ℓ$, as $χ(X) = (X-λ)(X-μ)$, with $λ≠μ$; we call
  this the \emph{Elkies case}.
\item $χ$ does not split modulo ℓ; we call this the \emph{Atkin case};
\item $χ$ is a square modulo $ℓ$.
\end{itemize}

The SEA algorithm, treats each of these cases in a slightly different
way; for simplicity, we will only sketch the Elkies case. %
In this case, there exists a basis $〈P,Q〉$ for $E[ℓ]$ onto which $π$
acts as a matrix
$M=\left(\begin{smallmatrix}λ&0\\0&μ\end{smallmatrix}\right)$. %
Each of the two eigenspaces of $M$ is the kernel of an isogeny of
degree $ℓ$ from $E$ to another curve $E'$. %
If we can determine the curve corresponding to, e.g., $〈P〉$, then we
can compute the isogeny $ϕ:E\to E/〈P〉$, and use it to formally
represent the point $P$. %
Then, $λ$ is recovered by solving the equation
\[[λ]P = π(P),\]
and from it we recover $t_π = λ + q/λ \mod \ell$.

Elkies' method is very similar to Schoof's original way of computing
$t_π$, however it is considerably more efficient thanks to the degree
of the extension rings involved. %
Indeed, in Schoof's algorithm a generic point of $E[ℓ]$ is represented
modulo the division polynomial $ψ_ℓ$, which has degree $(ℓ^2-1)/2$. %
In Elkies' algorithm, instead, the formal representation of $〈P〉$
only requires working modulo a polynomial of degree $≈ℓ$.

The other cases have similar complexity gains. %
For a more detailed overview, we address the reader
to~\cite{schoof95,lercier-algorithmique,elkies98,sutherland10}.

\section{Isogeny graphs}

We now look at the graph structure that isogenies create on the set of
$j$-invariants defined over a finite field. %
We start with an easy generalization of the Sato-Tate
theorem~\ref{th:sato-tate}.

\begin{theorem}[Sato-Tate]
  \label{th:sato-tate2}
  Two elliptic curves $E,E'$ defined over a finite field are isogenous
  if and only if their endomorphism algebras $\End(E)⊗ℚ$ and
  $\End(E')⊗ℚ$ are isomorphic.
\end{theorem}

An equivalence class of isogenous elliptic curves is called an
\emph{isogeny class}. %
In particular, we see that it is impossible for an isogeny class to
contain both ordinary and supersingular curves. %
When we restrict to isogenies of a prescribed degree $ℓ$, we say that
two curves are $ℓ$-isogenous; by the dual isogeny theorem, this too is
an equivalence relation. %
Remark that if $E$ is $ℓ$-isogenous to $E'$, and if $E''$ is
isomorphic to $E'$, then by composition $E$ and $E''$ are also
$ℓ$-isogenous. %

At this stage, we are only interested in elliptic curves up to
isomorphism, i.e., $j$-invariants. %
Accordingly, we say that two $j$-invariants are \emph{isogenous}
whenever their corresponding curves are. %

\begin{definition}[Isogeny graph]
  An \emph{isogeny graph} is a (multi)-graph which nodes are the
  $j$-invariants of isogenous curves, and which edges are isogenies
  between them.
\end{definition}

\begin{figure}
  \centering
    \begin{tikzpicture}
      \begin{scope}[xshift=6cm]
        \def\crater{7}
        \foreach \i in {1,...,\crater} {
          \draw[fill] (360/\crater*\i:1cm) circle (5pt);
          \draw (360/\crater*\i : 1cm) -- (360/\crater*\i+360/\crater : 1cm);
          \foreach \j in {-1,1} {
            \draw[fill] (360/\crater*\i : 1cm) -- (360/\crater*\i + \j*360/\crater/4 : 2cm) circle (3pt);
            \foreach \k in {-1,0,1} {
              \draw[fill] (360/\crater*\i + \j*360/\crater/4 : 2cm) --
              (360/\crater*\i + + \j*360/\crater/4 + \k*360/\crater/6 : 2.5cm) circle (1pt);
            }
          }
        }
      \end{scope}
      \begin{scope}[xshift=12cm]
        \node at (0,2) {$\End(E)$};
        \draw[fill] (0,1) circle(5pt) node[xshift=0.7cm]{$\O_K$} -- 
        (0,0) circle(3pt) --
        (0,-1) circle(1pt) node[xshift=0.7cm]{$ℤ[π]$};
      \end{scope}
    \end{tikzpicture}
  
    \caption{A volcano of $3$-isogenies (ordinary elliptic curves,
      Elkies case), and the corresponding tower of orders inside the
      endomorphism algebra.}
  \label{fig:volcano}
\end{figure}
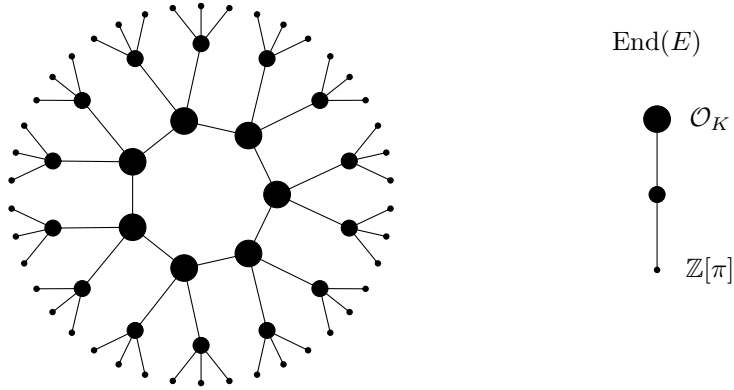

The dual isogeny theorem guarantees that for every isogeny $E\to E'$
there is a corresponding isogeny $E'\to E$ of the same degree. %
For this reason, isogeny graphs are usually drawn undirected. %
Figure~\ref{fig:volcano} shows a typical example of isogeny graph,
where we restrict to isogenies of degree $3$.

The classification of isogeny graphs was initiated by
Mestre~\cite{Mestre}, Pizer~\cite{pizer1,pizer2} and
Kohel~\cite{kohel}; further algorithmic treatment of graphs of
ordinary curves, and the now famous name of \emph{isogeny volcanoes}
was subsequently given by Fouquet and
Morain~\cite{fouquet+morain02}. %
We start with some generalities.

\begin{proposition}
  \label{th:isog-count}
  Let $E:y^2=x^3+ax+b$ be an elliptic curve defined over a finite
  field $k$ of characteristic $p$, and let $ℓ≠p$ be a prime.

  \begin{enumerate}
  \item There are $ℓ+1$ distinct isogenies of degree $ℓ$ with domain
    $E$ defined over the algebraic closure $\bar{k}$.
  \item There are $0,1,2$ or $ℓ+1$ isogenies of degree $ℓ$ with domain
    $E$ defined over $k$.
  \item If $E$ is ordinary, there is a unique separable isogeny of
    degree $p$ with domain $E$; there are none if $E$ is
    supersingular.
  \item The map $(x,y)\mapsto(x^p,y^p)$ is a purely inseparable
    isogeny of degree $p$ from $E$ to $E^{(p)}:y^2=x^3+a^px+b^p$.
  \end{enumerate}
\end{proposition}

There are many differences between the structure of isogeny graphs of
ordinary curves and those of supersingular ones. %
We focus here on the ordinary case, and we leave the supersingular one
for the last part.

\begin{proposition}[Horizontal and vertical isogenies]
  Let $ϕ:E\to E'$ be an isogeny of prime degree $ℓ$, and let $\O,\O'$
  be the orders corresponding to $E,E'$. %
  Then, either $\O⊂\O'$ or $\O'⊂\O$, and one of the following is true:
  \begin{itemize}
  \item $\O=\O'$, in this case $ϕ$ is said to \emph{horizontal};
  \item $[\O':\O]=ℓ$, in this case $ϕ$ is said to be \emph{ascending};
  \item $[\O:\O']=ℓ$, in this case $ϕ$ is said to be \emph{descending}.
  \end{itemize}
\end{proposition}
\begin{proof}
  See~\cite[Prop.~21]{kohel}.
\end{proof}

Observe that vertical isogenies can only exist for primes that divide
the conductor of $ℤ[π]$, so the horizontal case is the generic one. %
Like we did for the SEA algorithm we can further distinguish three
cases, depending on the value of the Legendre symbol
$\left(\frac{D}{ℓ}\right)$, i.e., depending on whether $π$ splits
(Elkies case), is inert (Atkin case), or ramifies modulo $ℓ$. %
All possible cases are encoded in the following proposition.

\begin{proposition}
  Let $E$ be an elliptic curve over a finite field $k$. %
  Let $\O$ be its endomorphism ring, $f$ its conductor, $D$ its
  discriminant, $π$ the Frobenius endormphism, $f_π$ the conductor of
  $ℤ[π]$. %
  Let $ℓ$ be a prime different from the characteristic of $k$, then
  the types of degree $ℓ$ isogenies with domain $E$ are as follows:
  \begin{itemize}
  \item If $ℓ|f$ and $ℓ\nmid(f_π/f)$, there is one ascending isogeny;
  \item If $ℓ|f$ and $ℓ|(f_π/f)$, there is one ascending isogeny and
    $ℓ$ descending ones;
  \item If $ℓ\nmid f$ and $ℓ\nmid(f_π/f)$, there are
    $1+\left(\frac{D}{ℓ}\right)$ horizontal isogenies, where
    $\left(\frac{D}{ℓ}\right)$ represents the Legendre symbol;
    \item If $ℓ\nmid f$ there are $1+\left(\frac{D}{ℓ}\right)$
      horizontal isogenies, plus $ℓ-\left(\frac{D}{ℓ}\right)$
      descending isogenies only if $ℓ|(f_π/f)$.
  \end{itemize}
\end{proposition}
\begin{proof}
  See~\cite[Prop.~21]{kohel}.
\end{proof}

Putting the pieces together, we see that graphs of ordinary curves
have a very rigid structure: a cycle of horizontal isogenies (Elkies
case), possibly reduced to one point (Atkin case), or to two points
(ramified case); and a tree of descending isogenies of height
$v_ℓ(f_π)$ (the $ℓ$-adic valuation of the conductor of $π$). %
Such graphs are called \emph{isogeny volcanoes} for obvious reasons
(have a look at Figure~\ref{fig:volcano}).

The action of $π$ on $E[ℓ]$, or more generally on $E[ℓ^k]$ for $k$
large enough, can be used to determine even more precisely which
isogenies are ascending, descending or horizontal. %
We will not give details here, but
see~\cite{MiretMRV05,MiretMSTV06,ionica+joux13,defeo2016explicit}.

\section{Application: computing irreducible polynomials }

In the applications seen in the first part, we have followed an old
\emph{mantra}: whenever an algorithm relies solely on the properties
of the multiplicative group $\F_q^*$, it can be generalized by
replacing $\F_q^*$ with the group of points of an elliptic curve over
$\F_q$ (or, eventually, a higher dimensional Abelian variety). %
Typically, the generalization adds some complexity to the computation,
but comes with the advantage of having more freedom in the choice of
the group size and structure. %
We now present another instance of the same \emph{mantra}, that is
particularly remarkable in our opinion: to the best of our knowledge,
it is the first algorithm where replacing $\F_q^*$ with $E(\F_q)$
required some non-trivial work with isogenies.

Constructing irreducible polynomials of arbitrary degree over a finite
field $\F_q$ is a classical problem. %
A classical solution consists in picking polynomials at random, and
applying an irreducibility test, until an irreducible one is found. %
This solution is not satisfactory for at least two reasons: it is not
deterministic, and has average complexity quadratic both in the degree
of the polynomial and in $\log q$.

For a few special cases, we have well known irreducible polynomials. %
For example, when $d$ divides $q-1$, there exist $α∈\F_q$ such that
$X^d-α$ is irreducible. %
Such an $α$ can be computed using Hilbert's theorem 90, or --more
pragmatically, and assuming that the factorization of $q-1$ is known--
by taking a random element and testing that it has no $d$-th root in
$\F_q$. %
It is evident that this algorithm relies on the fact that the
multiplicative group $\F_q^*$ is cyclic of order $q-1$.

At this point our \emph{mantra} suggests that we replace $α$ with a
point $P∈E(\F_q)$ that has no $ℓ$-divisor in $E(\F_q)$, for some well
chosen curve $E$. %
The obvious advantage is that we now require $ℓ|\#E(\F_q)$, thus we
are no longer limited to $ℓ|(q-1)$; however, what irreducible
polynomial shall we take? %
Intuition would suggest that we take the polynomial defining the
$ℓ$-divisors of $P$; however we know that the map $[ℓ]$ has degree
$ℓ^2$, thus the resulting polynomial would have degree too large, and
it would not even be irreducible.

This idea was first developed by Couveignes and
Lercier~\cite{couveignes+lercier11} and then slightly generalized
in~\cite{DeDoSc13}. %
Their answer to the question is to decompose the map $[ℓ]$ as a
composition of isogenies $\hat{ϕ}∘ϕ$, and then take the (irreducible)
polynomial vanishing on the fiber $ϕ^{-1}(P)$.

More precisely, let $\F_q$ be a finite field, and let $ℓ\nmid(q-1)$ be
odd and such that $ℓ\ll q+1+2\sqrt{q}$. %
Then there exists a curve $E$ which cardinality $\#E(\F_q)$ is
divisible by $\ell$. %
The hypothesis $ℓ\nmid(q-1)$ guarantees that $G = E[ℓ]∩E(\F_q)$ is
cyclic (see Exercice~\ref{ex:group-struct}). %
Let $ϕ$ be the degree $ℓ$ isogeny of kernel $G$, and let $E'$ be its
image curve. %
Let $P$ be a point in $E'(\F_q)\setminus [ℓ]E'(\F_q)$, Couveignes and
Lercier show that $\phi^{-1}(P)$ is an \emph{irreducible fiber}, i.e.,
that the polynomial
\[f(X) = \prod_{Q\in\phi^{-1}(P)}(X - x(Q))\]
is irreducible over $\F_q$.

To effectively compute the polynomial $f$, we need one last technical
ingredient: a way to compute a representation of the isogeny $ϕ$ as a
rational function. %
This is given to us by the famous V\'elu's formulas~\cite{velu71}.

\begin{proposition}[V\'elu's formulas]
  \label{th:velu}
  Let $E:y^2=x^3+ax+b$ be an elliptic curve defined over a field $k$,
  and let $G⊂E(\bar{k})$ be a finite subgroup. %
  The separable isogeny $ϕ:E\to E/G$, of kernel $G$, can be written as
  \begin{equation*}
    ϕ(P) = \left(
      x(P) + \sum_{Q∈G\setminus\{\O\}}x(P+Q)-x(Q),
      y(P) + \sum_{Q∈G\setminus\{\O\}}y(P+Q)-y(Q)
    \right);
  \end{equation*} %
  and the curve $E/G$ has equation $y^2=x^3+a'x+b'$, where
  \begin{align*}
    a' &= a - 5\sum_{Q∈G\setminus\{\O\}}(3x(Q)^2+a),\\
    b' &= b - 7\sum_{Q∈G\setminus\{\O\}}(5x(Q)^3+3ax(Q)+b).
  \end{align*}
\end{proposition}
\begin{proof}
  See~\cite[\S8.2]{df+thesis}.
\end{proof}

\begin{corollary}
  Let $E$ and $G$ be as above. %
  Let
  \[h(X) = \prod_{Q∈G\setminus\{\O\}}(X-x(Q)).\]
  Then the isogeny $ϕ$ can be expressed as
  \[ϕ(X,Y) = \left(\frac{g(X)}{h(X)}, y\left(\frac{g(x)}{h(x)}\right)'\right),\]
  where $g(X)$ is defined by
  \[\frac{g(X)}{h(X)} = dX-p_1-(3X^2+a)\frac{h'(X)}{h(X)}
    - 2(X^3+aX+b)\left(\frac{h'(X)}{h(X)}\right)',\]
  with $p_1$ the trace of $h(X)$ and $d$ its degree.
\end{corollary}
\begin{proof}
  See~\cite[\S8.2]{df+thesis}.
\end{proof}

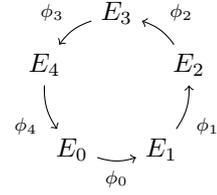
\begin{figure}
  \begin{subfigure}{0.65\textwidth}
    \begin{algorithmic}[1]
      \REQUIRE A finite field $\F_q$,\\
      a prime power $ℓ^e$ such that $ℓ\nmid(q-1)$ and $ℓ\ll q$;
      \ENSURE An irreducible polynomial of degree $ℓ^e$.
      \STATE Take random curves $E_0$, until one with $ℓ|\#E_0$ is found;
      \STATE Factor $\#E_0$;
      \FOR {$1≤i≤e$}
      \STATE Use V\'elu's formulas to compute a degree $ℓ$ isogeny $ϕ_i:E_{i-1}\to E_i$;
      \ENDFOR
      \STATE Take random points $P\in E_i(\F_q)$ until one not in $[ℓ]E_i(\F_q)$ is found;
      \RETURN The polynomial vanishing on the abscissas of $ϕ_i^{-1}∘\cdots∘ϕ_1^{-1}(P)$.
    \end{algorithmic}
  \end{subfigure}
  \hfill
  \begin{subfigure}{0.2\textwidth}
    \begin{tikzpicture}
      \def\n{4}
      \foreach \i in {0,...,\n} {
        \pgfmathparse{360/(\n+1)*(\i-1/2) - 90}
        \let\angle\pgfmathresult
        \draw (\angle:1) node (E\i) {$E_\i$};
      }
      \foreach \i in {0,...,\n} {
        \pgfmathparse{int(mod(\i+1, \n+1))}
        \let\j\pgfmathresult
        \draw (E\i) edge[->,bend right=18] node[auto,swap] {\scriptsize$ϕ_\i$} (E\j);
      }
    \end{tikzpicture}
  \end{subfigure}
  
  \caption{Couveignes-Lercier algorithm to compute irreducible
    polynomials, and structure of the computed isogeny cycle.}
  \label{fig:CL}
\end{figure}

The Couveignes-Lercier algorithm is summarized in
Figure~\ref{fig:CL}. %
What is most interesting, is the fact that it can be immediately
generalized to computing irreducible polynomials of degree $ℓ^e$, by
iterating the construction. %
Looking at the specific parameters, it is apparent that $ℓ$ is an
\emph{Elkies prime} for $E$ (i.e., $\left(\frac{D}{ℓ}\right)=1$), and
that each isogeny $ϕ_i$ is horizontal, thus their composition
eventually forms a cycle, the \emph{crater} of a volcano.

\section*{Exercices}

\begin{exercice}
  Prove Lemma~\ref{th:basis-change}.
\end{exercice}

\begin{exercice}
  Prove that $y$ divides the $m$-th division polynomial $ψ_m$ if and
  only if $m$ is even, and that no division polynomial is divisible by
  $y^2$.
\end{exercice}

\begin{exercice}
  Using the Sato-Tate theorem~\ref{th:sato-tate2}, prove that two
  curves are isogenous if and only if they have the same number of
  points.
\end{exercice}

\begin{exercice}
  Prove Propostion~\ref{th:isog-count}.
\end{exercice}

\begin{exercice}
  Prove that the dual of a horizontal isogeny is horizontal, and that
  the dual of a descending isogeny is ascending.
\end{exercice}

\begin{exercice}
  Prove that the height of a volcano of $ℓ$-isogenies is $v_ℓ(f_π)$,
  the $ℓ$-adic valuation of the Frobenius endomorphism.
\end{exercice}

\begin{exercice}
  Let $X^2-tX-q$ be the minimal polynomial of $π$, and suppose that it
  splits as $(X-λ)(X-μ)$ in $ℤ_ℓ$ (the ring of $ℓ$-adic integers). %
  Prove that the volcano of $ℓ$ isogenies has height $v_ℓ(λ-μ)$.
\end{exercice}

\begin{exercice}
  \label{ex:group-struct}
  Prove that $E[ℓ]⊂E(\F_q)$ implies $ℓ|(q-1)$.
\end{exercice}


\clearpage
\part{Cryptography from isogeny graphs }


\section{Expander graphs}

When we talk about \emph{Isogeny Based Cryptography}, as a topic
distinct from \emph{Elliptic Curve Cryptography}, we usually mean
algorithms and protocols that rely fundamentally on the structure of
\emph{large} isogeny graphs. %
The cryptographically interesting properties of these graphs are
usually tied to their \emph{expansion} properties. %

We recall some basic concepts of graph theory; for simplicity, we will
restrict to undirected graphs. %
An undirected graph $G$ is a pair $(V,E)$ where $V$ is a finite set of
\emph{vertices} and $E⊂V×V$ is a set of unordered pairs called
\emph{edges}. %
Two vertices $v,v'$ are said to be \emph{connected by an edge} if
$\{v,v'\}∈E$. %
The \emph{neighbors} of a vertex $v$ are the vertices of $V$ connected
to it by an edge. %
A \emph{path} between two vertices $v,v'$ is a sequence of vertices
$v\to v_1\to\cdots\to v'$ such that each vertex is connected to the
next by an edge. %
The \emph{distance} between two vertices is the length of the shortest
path between them; if there is no such path, the vertices are said to
be at infinite distance. %
A graph is called \emph{connected} if any two vertices have a path
connecting them; it is called \emph{disconnected} otherwise. %
The \emph{diameter} of a connected graph is the largest of all
distances between its vertices. %
The \emph{degree} of a vertex is the number of edges pointing to (or
from) it; a graph where every edge has degree $k$ is called
\emph{$k$-regular}. %
The \emph{adjacency matrix} of a graph $G$ with vertex set
$V=\{v_1,\dots,v_n\}$ and edge set $E$, is the $n×n$ matrix where the
$(i,j)$-th entry is $1$ if there is an edge between $v_i$ and $v_j$,
and $0$ otherwise. %
Because our graphs are undirected, the adjacency matrix is symmetric,
thus it has $n$ real eigenvalues
\[λ_1≥\cdots≥λ_n.\] %
It is convenient to identify functions on $V$ with vectors in $ℝ^n$,
and therefore also think of the adjacency matrix as a self-adjoint
operator on $L^2(V)$. %
Then can we immediately bound the eigenvalues of $G$.

\begin{proposition}
  \label{th:graph-eigen}
  If $G$ is a $k$-regular graph, then its largest and smallest
  eigenvalues $λ_1,λ_n$ satisfy
  \[k=λ_1≥λ_n≥-k.\]
\end{proposition}
\begin{proof}
  See~\cite[Lem.~2]{tao2011expander}.
\end{proof}

\begin{definition}[Expander graph]
  Let $ε>0$ and $k≥1$. A $k$-regular graph is called a (one-sided)
  \emph{$ε$-expander} if
  \[λ_2≤(1-ε)k;\]
  and a \emph{two-sided $ε$-expander} if it also satisfies
  \[λ_n≥-(1-ε)k.\] %
  A sequence $G_i=(V_i,E_i)$ of $k$-regular graphs with $\#V_i\to∞$ is
  said to be a one-sided (resp. two-sided) \emph{expander family} if
  there is an $ε>0$ such that $G_i$ is a one-sided (resp. two-sided)
  $ε$-expander for all sufficiently large $i$.
\end{definition}

\begin{theorem}[Ramanujan graph]
  Let $k≥1$, and let $G_i$ be a sequence of $k$-regular graphs. %
  Then
  \[\max(|λ_2|,|λ_n|) ≥ 2\sqrt{k-1} - o(1),\]
  as $n\to ∞$. %
  A graph such that $|λ_i|≤2\sqrt{k-1}$ for any $λ_i$ except $λ_1$ is
  called a \emph{Ramanujan graph}.
\end{theorem}

The \emph{spectral} definition of expansion is very practical to work
with, but gives very little intuition on the topological properties of
the graph. %
\emph{Edge expansion} quantifies how well subsets of vertices are
connected to the whole graph, or, said otherwise, how far the graph is
from being disconnected.

\begin{definition}[Edge expansion]
  Let $F⊂V$ be a subset of the vertices of $G$. %
  The \emph{boundary of $F$}, denoted by $∂F⊂E$, is the subset of the
  edges of $G$ that go from $F$ to $V\setminus F$. %
  The \emph{edge expansion ratio} of $G$, denoted by $h(G)$ is the
  quantity
  \[h(G) = \min_{\substack{F⊂V,\\ \#F≤\#V/2}}\frac{\#∂F}{\#F}.\]
\end{definition}

Note that $h(G)=0$ if and only if $G$ is disconnected. %
Edge expansion is strongly tied to spectral expansion, as the
following theorem shows.

\begin{theorem}[Discrete Cheeger inequality]
  Let $G$ be a $k$-regular one-sided $ε$-expander, then
  \[\frac{ε}{2}k≤h(G)≤\sqrt{2ε}k.\]
\end{theorem}

Expander families of graphs have many applications in theoretical
computer science, thanks to their \emph{pseudo-randomness} properties:
they are useful to construct \emph{pseudo-random number generators},
\emph{error-correcting codes}, \emph{probabilistic checkable proofs},
and, most interesting to us, \emph{cryptographic primitives}. %
Qualitatively, we can describe them as having \emph{short diameter}
and \emph{rapidly mixing walks}.

\begin{proposition}
  Let $G$ be a $k$-regular one sided $ε$-expander. %
  for any vertex $v$ and any radius $r>0$, let $B(v,r)$ be the
  \emph{ball} of vertices at distance at most $r$ from $v$. %
  Then, there is a constant $c>0$, depending only on $k$ and $ε$, such
  that
  \[\#B(v,r)≥\min((1+c)^r,\#V).\]
\end{proposition}

In particular, this shows that the diameter of an expander is bounded
by $O(\log n)$, where the constant depends only on $k$ and $ε$. %
A \emph{random walk} of length $i$ is a path $v_1\to\cdots\to v_i$,
defined by the random process that selects $v_i$ uniformly at random
among the neighbors of $v_{i-1}$. %
Loosely speaking, the next proposition says that, in an expander
graph, random walks of length close to its diameter terminate on any
vertex with probability close to uniform. %

\begin{proposition}[Mixing theorem]
  Let $G=(V,E)$ be a $k$-regular two-sided $ε$-expander. %
  Let $F⊂V$ be any subset of the vertices of $G$, and let $v$ be any
  vertex in $V$. %
  Then a random walk of length at least
  \[\frac{\log\#F^{1/2}/2\#V}{\log(1-ε)}\] %
  starting from $v$ will land in $F$ with probability at least
  $\#F/2\#V$.
\end{proposition}
\begin{proof}
  See~\cite{JMV}.
\end{proof}

The length in the previous proposition is also called the \emph{mixing
  length} of the expander graph. %
We conclude this section with two results on expansion in graphs of
isogenies.

\begin{theorem}[Supersingular graphs are Ramanujan]
  \label{th:ss-exp}
  Let $p,\ell$ be distinct primes, then
  \begin{enumerate}
  \item All supersingular $j$-invariants of curves in $\bar\F_p$ are
    defined in $\F_{p^2}$;
  \item There are
    \begin{equation*}
      \lfloor\frac{p}{12}\rfloor +
      \begin{cases}
        0 &\text{if $p=1\mod 12$}\\
        1 &\text{if $p=5,7\mod 12$}\\
        2 &\text{if $p=11\mod 12$}
      \end{cases}
    \end{equation*}
    isomorphism classes of supersingular elliptic curves over
    $\bar\F_p$;
  \item The graph of supersingular curves in $\bar\F_p$ with
    $ℓ$-isogenies is connected, $ℓ+1$ regular, and has the Ramanujan
    property.
  \end{enumerate}
\end{theorem}
\begin{proof}
  See~\cite[V, Th.~4.1]{silverman:elliptic}, \cite{pizer1,pizer2},
  \cite{charles+lauter+goren09}.
\end{proof}

\begin{theorem}[Graphs of horizontal isogenies are expanders]
  \label{th:ord-exp}
  Let $\F_q$ be a finite field and let $\O⊂ℚ[\sqrt{-D}]$ be an order
  in a quadratic imaginary field. %
  Let $G$ be the graph which vertices are elliptic curves over $\F_q$
  with complex multiplication by $\O$, and which edges are
  (horizontal) isogenies of prime degree bounded by $(\log q)^{2+δ}$
  for some fixed $δ>0$. %
  Assume that $G$ is non-empty. %
  Then, under the generalized Riemann hypothesis, $G$ is a regular
  graph and there exists an $ε$, independent of $\O$ and $q$, such
  that $G$ is a one-sided $ε$-expander.
\end{theorem}
\begin{proof}
  See~\cite{JMV}.
\end{proof}

\section{Isogeny graphs in cryptanalysis}
\label{sec:isog-graphs-crypt}

Besides the applications to point counting mentioned in the previous
part, the first application of isogenies in cryptography has been to
study the difficulty of the discrete logarithm problem in elliptic
curves. %
One can state several computational problems related to isogenies,
both \emph{easy} and \emph{hard} ones. %
Here are some examples.

\begin{problem}[Isogeny computation]
  Given an elliptic curve $E$ with Frobenius endomorphism $π$, and a
  subgroup $G⊂E$ such that $π(G)=G$, compute the rational fractions
  and the image curve of the separable isogeny $ϕ$ of kernel $G$.
\end{problem}

V\'elu's formulas (Proposition~\ref{th:velu}) give a solution to this
problem in $\tildO(\#G)$ operations over the field of definition of
$E$. %
This is nearly optimal, given that the output has size $O(\#G)$.
  
However in some special instances, e.g., when $ϕ$ is a composition of
many small degree isogenies, the rational fractions may be represented
more compactly, and the cost may become only logarithmic in $\#G$.

\begin{problem}[Explicit isogeny]
  \label{prob:explicit-isog}
  Given two elliptic curves $E,E'$ over a finite field, isogenous
  of known degree $d$, find an isogeny $ϕ:E\to E'$ of degree $d$.
\end{problem}

Remark that, up to automorphisms, the isogeny $ϕ$ is typically
unique. %
Elkies was the first to formulate the problem and give an
algorithm~\cite{elkies92,elkies98} with complexity $O(d^3)$ in
general, and $\tildO(d)$ in the special context of the SEA
algorithm~\cite{bostan+morain+salvy+schost08,lercier+sirvent08}. %
Alternate algorithms, with complexity $O(d^2)$ in general, are due to
Couveignes and
others~\cite{couveignes94,couveignes96,couveignes00,df+schost09,defeo2016explicit}.

\begin{problem}[Isogeny path]
  Given two elliptic curves $E,E'$ over a finite field $k$, such that
  $\#E=\#E'$, find an isogeny $ϕ:E\to E'$ of smooth degree.
\end{problem}

This problem, and variations thereof, is the one that occurs most in
isogeny based cryptography. %
It is a notoriously difficult problem, for which only algorithms
exponential in $\log\#E$ are known in general. %
A general strategy to tackle it is by a \emph{meet in the middle}
random walk~\cite{galbraith99}: choose an expander graph $G$
containing both $E$ and $E'$, and start a random walk from each
curve. %
By the birthday paradox, the two walks are expected to meet after
roughly $O(\sqrt{\#G})$ steps; when a collision is detected, the 
composition of the walks yields the desired isogeny.

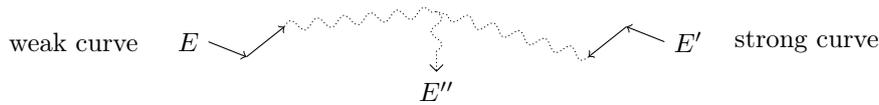
\begin{figure}
  \centering
  \begin{tikzpicture}
    \path (0,0) node[anchor=east] {$E$} (6,0) node[anchor=west] {$E'$};
    \path (-0.8,0) node[anchor=east] {weak curve}
    (6.8,0) node[anchor=west] {strong curve};

    \draw[->] (0,0) -- (0.5,-0.2);
    \draw[->] (6,0) -- (5.5,0.2);
    \draw[->] (0.5,-0.2) -- (1,0.2);
    \draw[->] (5.5,0.2) -- (5,-0.2);
    \begin{scope}[densely dotted,coils/.style={decorate,decoration={coil,aspect=0,amplitude=2pt}}]
      \draw[coils] (1,0.2) -- (3,0.4);
      \draw[coils] (5,-0.2) -- (3,0.4);
      \draw[-angle 90,coils] (3,0.4) -- (3, -0.4) node[anchor=north] {$E''$};
    \end{scope}
  \end{tikzpicture}
  \caption{The meet in the middle attack in weak isogeny classes.}
  \label{fig:ghs}
\end{figure}

The meet in the middle strategy was notoriously used to extend the
power of the GHS attack on elliptic curves defined over extension
fields of composite degree~\cite{gaudry+hess+smart02,GHS}. %
Without going into the details of the GHS attacks, one of its
remarkable properties is that only a small fraction of a given isogeny
class is vulnerable to it. %
Finding an isogeny from an immune curve to a weak curve allows the
attacker to map the discrete logarithm problem from one to the
other. %
The average size of an isogeny class of ordinary elliptic curves is
$O(\sqrt{\#E})$, thus the meet in the middle strategy yields an
$O(\#E^{1/4})$ attack on any curve in the class: better than a generic
attack on the discrete logarithm problem. %
The attack is pictured in Figure~\ref{fig:ghs}.

Similar ideas have been used to construct \emph{key escrow
  systems}~\cite{teske06}, and to prove random reducibility of
discrete logarithms inside some isogeny classes~\cite{JMV}.

\section{Provably secure hash functions}

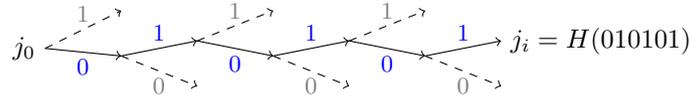
\begin{figure}
  \centering
  \begin{tikzpicture}
    \coordinate (last) at (0,0);
    \draw (last) node[anchor=east] {$j_0$};
    \foreach \i in {1,...,6} {
      \pgfmathparse{(-1)^\i}
      \let\sign\pgfmathresult
      \pgfmathparse{int(mod(\i+1,2))}
      \let\bit\pgfmathresult
      \pgfmathparse{int(mod(\i,2))}
      \let\nbit\pgfmathresult
      \draw[->] (last) -- (\i,\sign*0.1) node[blue,pos=0.5,yshift=\sign*0.2cm]{\small$\bit$};
      \draw[dashed,->] (last) -- (\i,-\sign*0.5) node[gray,pos=0.5,yshift=-\sign*0.2cm]{\small$\nbit$};
      \coordinate (last) at (\i,\sign*0.1);
    }
    \draw (last) node[anchor=west] {$j_i=H(010101)$};
  \end{tikzpicture}
  \caption{Hashing the string $010101$ using an expander graph}
  \label{fig:hash}
\end{figure}

The next application of isogeny graphs is constructing \emph{provably
  secure hash functions}. %
The mixing properties of expander graphs make them very good
pseudo-random generators. %
For the very same reason, they can also be used to define \emph{hash
  functions}. %
The Charles-Goren-Lauter (CGL)
construction~\cite{charles+lauter+goren09} chooses an arbitrary start
vertex $j_0$ in an expander graph, then takes a random walk (without
backtracking) according to the string to be hashed, and outputs the
arrival vertex. %
To fix notation, let's assume that the graph is $3$-regular, then the
value to be hashed is encoded as a binary string. %
At each step one bit is read from the string, and its value is used to
choose an edge from the current vertex to the next one, avoiding the
one edge that goes back. %
The way an edge is chosen according to the read bit need only be
deterministic, but can be otherwise arbitrary (e.g., determined by
some lexicographic ordering). %
The process is pictured in Figure~\ref{fig:hash}.

For the process to be a good pseudo-random function, the walks need to
be longer than the mixing length of the graph. %
However this is not enough to guarantee a \emph{cryptographically
  strong} hash function. %
Indeed the two main properties of cryptographic hash functions,
translate in this setting as the following computational problems.

\begin{problem}[Preimage resistance]
  Given a vertex $j$ in the graph, find a path from the start vertex
  $j_0$ to $j$.
\end{problem}

\begin{problem}[Collision resistance]
  Find a non-trivial loop (i.e., one that does not track backwards)
  from $j_0$ to itself.
\end{problem}

Charles, Goren and Lauter suggested two types of expander graphs to be
used in their constructions. %
One is based on \emph{Cayley graphs}, and was broken shortly
afterwards~\cite{tillich2008collisions,quis}. %
The second one is based on graphs of supersingular curves. %
In this context, the preimage finding problem is an instance of the
isogeny path problem, while the collision finding problem is
equivalent to computing a non-trivial endomorphism of the start curve
$j_0$. %
In this sense, the CGL hash function on expander graphs has
\emph{provable security}, meaning that its cryptographic strength can
be provably reduced to well defined mathematical problems thought to
be hard.

Nevertheless, the CGL hash function has failed to attract the interest
of practitioners. %
For one, it is considerably slower than popular hash functions such as
those standardized by NIST. %
More worryingly, some weaknesses have recently been
highlighted~\cite{kohel2014quaternion,cryptoeprint:2017:962}, that
could potentially lead to backdoors in standardized parameters.

\section{Post-quantum key exchange}
\label{sec:post-quantum-key}

We come to the last, more powerful constructions based on isogeny
graphs. %
We present here two key exchange protocols, similar in spirit to the
Diffie-Hellman protocol discussed in the first part. %
Both protocols are significantly less efficient than ECDH, however
they are relevant because of their conjectured \emph{quantum
  security}. %
In recent years, the case has been made that cryptographic standards
must be amended, in view of the potential threat of general purpose
\emph{quantum computers} becoming available. %
It is well known, indeed, that Shor's
algorithm~\cite{shor1994algorithms} would solve the factorization and
the discrete logarithm problems in polynomial time on a quantum
computer, thus sealing the fate of RSA, ECDH, and any other protocol
based on them. %
For this reason, the cryptographic community is actively seeking
cryptographic primitives that would not break in polynomial time on
quantum computers.

Both protocols are based on random walks in an isogeny graph. %
The two participants, Alice and Bob, start from the same common curve
$E_0$, and take a (secret) random walk to some curves $E_A,E_B$. %
After publishing their respective curves, Alice starts a new walk from
$E_B$, while Bob starts from $E_A$. %
By repeating the ``same'' secret steps, they both eventually arrive on
a shared secret curve $E_S$, only known to them. %
While the idea may seem simple, its realization is far from easy. %
Indeed, as opposed to the hash function case, we cannot be content
with an arbitrary labeling of the graph edges. %
We must instead use the algebraic properties of the isogeny graphs to
ensure that Alice and Bob's walks ``commute''.

\subsection{Hard homogeneous spaces}

The first protocol originates in a preprint by Couveignes~\cite{Couv},
but was only later put into practice and popularized by Rostovtsev and
Stolbunov~\cite{R&S,Stol}. %
It uses random walks in graphs of ordinary curves with horizontal
isogenies; in this sense, it is a direct application of
Theorem~\ref{th:ord-exp}. %
The protocol can be viewed as a special instance of a general
construction on \emph{Schreier graphs}, a generalization of
\emph{Cayley graphs}.

\begin{definition}[Schreier graph]
  Let $G$ be a group \emph{acting freely} on a set $X$, in the sense
  that there is a map
  \begin{align*}
    G×X &\to X\\
    (σ,x) &\mapsto σ·x
  \end{align*}
  such that $σ·x=x$ if and only if $σ=1$, and $σ·(τ·x)=(στ)·x$, for
  all $σ,τ∈G$ and $x∈X$. %
  Let $S⊂G$ be a \emph{symmetric} subset, i.e. one not containing $1$
  and closed under inversion. %
  The \emph{Schreier graph} of $(S,X)$ is the graph which vertices are
  the elements of $X$, and such that $x,x'∈X$ are connected by an edge
  if and only if $σ·x=x'$ for some $σ∈S$.
\end{definition}

Because of the constraints on the group action and the set $S$,
Schreier graphs are undirected and regular, and they usually make good
expanders (see exercise \ref{ex:schreier}). %
Note that Cayley graphs are the Schreier graphs of the (left) action
of a group on itself.

\begin{figure}
  \centering
  \begin{tikzpicture}
    \begin{scope}
      \def\crater{12}
      \def\jumpa{-8}
      \def\jumpb{9}
      \def\diam{2.5cm}

      \foreach \i in {1,...,\crater} {
        \draw[blue] (360/\crater*\i : \diam) to[bend right] (360/\crater*\i+360/\crater : \diam);
        \draw[red] (360/\crater*\i : \diam) to[bend right] (360/\crater*\i+\jumpa*360/\crater : \diam);
        \draw[green] (360/\crater*\i : \diam) to[bend right=50] (360/\crater*\i+\jumpb*360/\crater : \diam);
      }
      \foreach \i in {1,...,\crater} {
        \pgfmathparse{int(mod(2^\i,13))}
        \let\exp\pgfmathresult
        \draw[fill] (360/\crater*\i: \diam) circle (2pt) +(360/\crater*\i: 0.4) node{$g^{\exp}$};
      }
    \end{scope}
    \begin{scope}[xshift=4cm,yshift=1cm]
      \draw[blue] (0,0) -- (0.5,0) (0.5,0) node[black,anchor=west] {$x \mapsto x^{2}$};
      \draw[red] (0,-1) -- (0.5,-1) (0.5,-1) node[black,anchor=west] {$x \mapsto x^{3}$};
      \draw[green] (0,-2) -- (0.5,-2) (0.5,-2) node[black,anchor=west] {$x \mapsto x^{5}$};
    \end{scope}
  \end{tikzpicture}
  \caption{Schreier graph of the generators of a group of order $13$
    under the action of
    $S=\{2,3,5,2^{-1},3^{-1},5^{-1}\}⊂(ℤ/13ℤ)^{×}$.}
  \label{fig:schreier}
\end{figure}
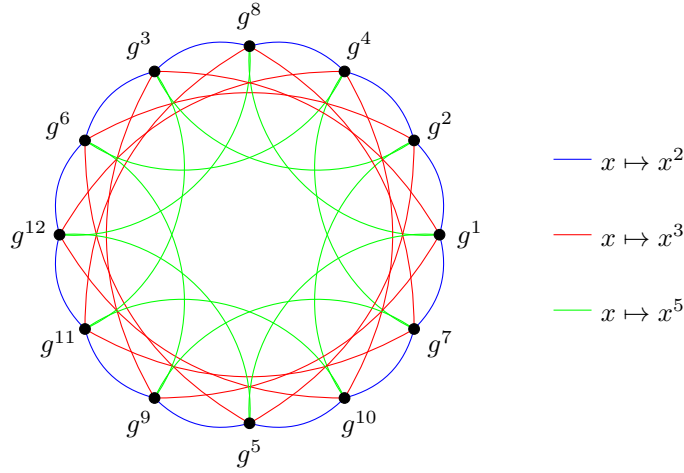

As an example, take a cyclic group $G$ of order $n$, then $(ℤ/nℤ)^{×}$
acts naturally on $G$ by the law $σ·g=g^σ$ for any $g∈G$ and
$σ∈(ℤ/nℤ)^{×}$. %
This action is not free on $G$, but it is so on the subset $P$ of all
generators of $G$; we can thus build the Schreier graph $(S,P)$, where
$S$ is a symmetric subset that generates $(ℤ/nℤ)^{×}$. %
An example of such graph for the case $n=13$ is shown in
Figure~\ref{fig:schreier}, where the set $S⊂(ℤ/13ℤ)^{×}$ has been
chosen to contain $2,3,5$ and their inverses.

By slightly generalizing Couveignes' work~\cite{Couv}, we will now
show how to construct a key exchange protocol based on this family of
Schreier graphs. %
We will restrict to cyclic groups of prime order $p$, and we will have
the cryptosystem security grow exponentially in $\log p$. %
Let $G=〈g〉$ be a cyclic group of order $p$; let $D⊂(ℤ/pℤ)^{×}$ be a
generating set such that $σ∈D$ implies $σ^{-1}∉D$; and let
$S = D∪D^{-1}$. %
We call \emph{directed route} a sequence of elements of $D$. %
A directed route $ρ∈D^*$, together with a \emph{starting vertex}
$g∈G$, defines a walk in the Schreier graph $(S,G)$ by starting in
$g$, and successively taking the edges corresponding to the labels in
$ρ$. %
If $ρ$ is a directed route, and $g∈G$, we write $ρ(g)$ for vertex
where the walk defined by $ρ$ and $g$ ends. %
We can now define a key exchange protocol where the secrets are random
directed routes, and the public data are vertices of the Schreier
graph. %
The protocol is summarized in Figure~\ref{fig:walk-dh}.

\begin{figure}
  \centering
  \begin{tabular}{l *{2}{p{30ex}<{\centering}}}
    \hline
    Public parameters & \multicolumn{2}{l}{A group $G$ of prime order $p$,}\\
                      & \multicolumn{2}{l}{A generating set $D⊂(ℤ/pℤ)^{×}$ such that $σ∈D⇒σ^{-1}∉D$,}\\
                      & \multicolumn{2}{l}{A generator $g$ of $G$.}\\
    \hline
                      & {\bf Alice} & {\bf Bob}\\
    \hline
    Pick random secret & $ρ_A∈D^*$ & $ρ_B∈D^*$\\
    Compute public data & $g_A = ρ_A(g)$ & $g_B = ρ_B(g)$\\
    Exchange data &  \hfill $g_A \longrightarrow$ & $\longleftarrow g_B$ \hfill\strut \\
    Compute shared secret & $g_{AB} = ρ_A(g_B)$ & $g_{AB} = ρ_B(g_A)$
  \end{tabular}
  
  \caption{Key exchange protocol based on random walks in a Schreier graph.}
  \label{fig:walk-dh}
\end{figure}

\begin{figure}
  \centering
  \newcommand{\myedge}[3]{
    \draw[#3] (360/\crater*#1 : \diam) to[bend right] (360/\crater*#2 : \diam);
  }
  \begin{tikzpicture}
    \begin{scope}
      \def\crater{12}
      \def\jumpa{-8}
      \def\jumpb{9}
      \def\diam{2cm}
      \foreach \i in {1,...,\crater} {
        \pgfmathparse{int(mod(2^\i,13))}
        \let\exp\pgfmathresult
        \draw[fill] (360/\crater*\i: \diam) circle (2pt);
      }
      \myedge{0}{1}{blue}\myedge{1}{5}{red}\myedge{5}{6}{blue}\myedge{6}{3}{green}
      \begin{scope}[dashed,thick]
        \myedge{3}{7}{red}\myedge{7}{11}{red}\myedge{11}{8}{green}\myedge{8}{9}{blue}
      \end{scope}
      \draw (0 : \diam + 0.4cm) node {$g$};
      \draw (360/\crater*3 : \diam + 0.4cm) node {$g_A$};
      \draw (360/\crater*9 : \diam + 0.4cm) node {$g_{AB}$};
    \end{scope}
    
    \begin{scope}[xshift=6.5cm]
      \def\crater{12}
      \def\jumpa{-8}
      \def\jumpb{9}
      \def\diam{2cm}
      \foreach \i in {1,...,\crater} {
        \pgfmathparse{int(mod(2^\i,13))}
        \let\exp\pgfmathresult
        \draw[fill] (360/\crater*\i: \diam) circle (2pt);
      }
      \begin{scope}[dashed,thick]
        \myedge{0}{4}{red}\myedge{4}{8}{red}\myedge{8}{5}{green}\myedge{5}{6}{blue}
      \end{scope}
      \myedge{6}{7}{blue}\myedge{7}{11}{red}\myedge{11}{0}{blue}\myedge{0}{9}{green}
      \draw (0 : \diam + 0.4cm) node {$g$};
      \draw (360/\crater*6 : \diam + 0.4cm) node {$g_B$};
      \draw (360/\crater*9 : \diam + 0.4cm) node {$g_{AB}$};
    \end{scope}
  \end{tikzpicture}  

  \caption{Example of key exchange on the Schreier graph of
    Figure~\ref{fig:schreier}.  Alice's route is represented by
    continuous lines, Bob's route by dashed lines. On the left, Bob
    computes the shared secret starting from Alice's public data. On
    the right, Alice does the analogous computation.}
  \label{fig:dh-walk-pict}
\end{figure}
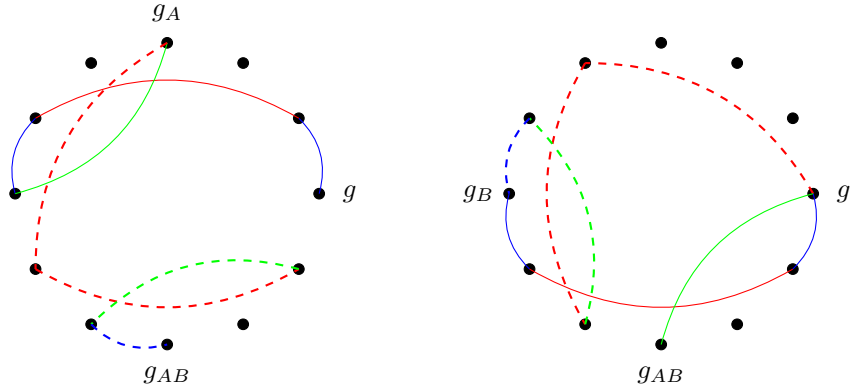

A graphical example of this protocol with $p=13$ and $D=\{2,3,5\}$ is
given in Figure~\ref{fig:dh-walk-pict}. %
To understand why it works, observe that if $ρ$ is a route of length $m$
\[ρ=(σ_1,\dots,σ_m),\]
then
\[ρ(g) = \exp_g\left(\prod σ_i\right)\] %
for any $g∈G$. %
Hence, the order of the steps in a route does not matter: what counts
is only how many times each element of $D$ appears in $ρ$. %
We immediately realize that this protocol is nothing else than the
classical Diffie-Hellman protocol on the group $G$, presented in a
twisted way.%
\footnote{A minor difference lies in the fact that this protocol
  avoids non-primitive elements of $G$, whereas the classical
  Diffie-Hellman protocol may well use public keys belonging to a
  subgroup of $G$.} %

For this protocol to have the same security as the original
Diffie-Helman, we need the public keys $g_A,g_B$ to be (almost)
uniformly distributed. %
Hence, we shall require that the graph is an expander, and that walks
are longer than the mixing length; i.e., that $D$ generates
$(ℤ/pℤ)^{×}$, and that walks have length $\sim\log p$. %
Since a secret route is simply defined by the number of times each
element of $D$ is present, we shall also need
$\#D\sim \log p/\loglog p$ in order to have a large enough key
space. %
If we respect all these constraints, we end up with a protocol that is
essentially equivalent to the original Diffie-Hellman, only less
efficient.

It is now an easy exercise to generalize to other Schreier graphs. %
To see how this applies to isogeny graphs, we must take a step back,
and define some more objects related to elliptic curves.

\begin{definition}[Fractional ideal]
  Let $\O$ be an order in a number field $K$. A \emph{fractional
    ideal} of $\O$ is a non-zero subgroup $I⊂K$ such that
  \begin{itemize}
  \item $xI⊂I$ for all $x∈\O$, and
  \item there exists a non-zero $x∈\O$ such that $xI⊂\O$.
  \end{itemize}
  
  A fractional ideal is called \emph{principal} if it is of the form
  $x\O$ for some $x∈K$.
\end{definition}

Note that the ideals of $\O$ are exactly the fractional ideals
contained in $\O$; however, from now on we will simply call
\emph{ideals} the fraction ideals, and we will use the name
\emph{integral ideal} for ordinary ones. %
An ideal $I$ is said to be \emph{invertible} if there is another ideal
$J$ such that $IJ=\O$. %
Invertible ideals form an Abelian group, written multiplicatively,
under the operation
\[IJ = \{ xy \;|\; x∈I, y∈J \}.\] %
It is easily verified that $\O$ is the neutral element of the group,
and that principal ideals form a subgroup of it.

\begin{proposition}[Ideal class group]
  Let $\O$ be an order in a number field $K$. %
  Let $\mathcal{I}(\O)$ be its group of invertible ideals, and
  $\mathcal{P}(\O)$ the subgroup of principal ideals. %
  The \emph{(ideal) class group} of $\O$ is the quotient
  \[\Cl(\O) = \mathcal{I}(\O)/\mathcal{P}(\O).\]
  It is a finite Abelian group. %
  Its order, denoted by $h(\O)$, is called the \emph{class number} of
  $\O$.
\end{proposition}

The class group is a fundamental object in the study of number fields
and their Galois theory. %
What is relevant to us, is the fact that the elements of $\Cl(\O)$ are
\emph{represented} by horizontal isogenies, a fact that is developed
in the theory of \emph{complex multiplication}. %
We only take here a small peek at the theory;
see~\cite{lang1987elliptic,silverman:advanced,cox2011primes} for a
detailed account.

\begin{definition}[$\a$-torsion]
  Let $E$ be an elliptic curve defined over a finite field $\F_q$. %
  Let $\O$ be the endomorphism ring of $E$, and let $\a⊂\O$
  be an integral invertible ideal of norm coprime to $q$. %
  We define the \emph{$\a$-torsion subgroup} of $E$ as
  \[E[\a] = \{P∈E \;|\; α(P) = 0 \text{ for all } α∈\a\}.\] %
\end{definition}

Given an ideal $\a⊂\O$ as above, it is natural to define the
(separable) isogeny $ϕ_{\a}:E\to E_{\a}$, where $E_{\a}=E/E[\a]$. %
This definition can be readily extended to inseparable isogenies. %
Since $\a$ is invertible, we can show that $\End(E)≃\End(E_\a)≃\O$,
that $E_\a$ only depends on the class of $\a$ in $\Cl(\O)$, and that
the map $(\a,E)\mapsto E_\a$ defines a group action of $\Cl(\O)$ on
the set of elliptic curves with complex multiplication by $\O$.

\begin{theorem}
  Let $\F_q$ be a finite field, and let $\O⊂ℚ[\sqrt{-D}]$ be an order
  in a quadratic imaginary field. %
  Denote by $\Ell_q(\O)$ the set of elliptic curves defined over
  $\F_q$ with complex multiplication by $\O$.

  Assume that $\Ell_q(\O)$ is non-empty, then the class group
  $\Cl(\O)$ acts \emph{freely} and \emph{transitively} on it; i.e.,
  there is a map
  \begin{align*}
    \Cl(\O)×\Ell_q(\O) &\to \Ell_q(\O)\\
    (\a,E) &\mapsto \a·E
  \end{align*}
  such that $\a·(\mathfrak{b}·E)=(\a\mathfrak{b})·E$ for all
  $\a,\mathfrak{b}∈\Cl(\O)$ and $E∈\Ell_q(\O)$, and such that for any
  $E,E'∈\Ell_q(\O)$ there is a unique $\a∈\Cl(\O)$ such that
  $E'=\a·E$. %
\end{theorem}

A set that is acted upon freely and transitively by a group $G$, is
also called a \emph{principal homogeneous space} or a \emph{torsor}
for $G$. %
An immediate consequence of the theorem above is that the torsor
$\Ell_q(\O)$ has cardinality equal to the class number $h(\O)$. %

Following on from the connection between isogenies and ideals, suppose
that that $ℓ\O$ splits into prime ideals as
$ℓ\O=\mathfrak{l}\bar{\mathfrak{l}}$. %
Set $S=\{\mathfrak{l},\bar{\mathfrak{l}}\}$, then the Schreier graph
of $(S, \Ell_q(\O))$ is exactly the graph of horizontal
$\ell$-isogenies on $\Ell_q(\O)$. %
More generally, if we let $S⊂\Cl(\O)$ be a symmetric subset, its
Schreier graph is a graph of horizontal isogenies, and it is an
expander if and only if $S$ generates $\Cl(\O)$.

Based on this observation, we can now give a key exchange protocol
based on random walks in graphs of horizontal isogenies. %
The general idea was already present in Couveignes' work~\cite{Couv},
but it was Rostovtsev and Stolbunov who proposed to use isogeny
computations to effectively implement the protocol~\cite{R&S,Stol}. %

The protocol implicitly uses the set $\Ell_q(\O)$ of elliptic curves
over $\F_q$ with complex multiplication by some order $\O$; however it
never explicitly computes $\O$. %
Instead, it determines parameters in the following order:
\begin{enumerate}
\item A \emph{large enough} finite field $\F_q$;
\item A curve $E$ defined over $\F_q$;
\item The Frobenius discriminant $D_π=t_π^2-4q$ of $E$ is computed
  through point counting, and it is verified that it contains a
  \emph{large enough} prime factor;
\item A set $L=\{ℓ_1,\dots,ℓ_m\}$ of primes that split in $ℤ[π]$,
  i.e., such that $\left(\frac{D_π}{ℓ_i}\right)=1$;
\item For each prime $ℓ_i$, the factorization
  \[π^2-t_ππ+q=(π-λ_i)(π-μ_i)\mod ℓ_i\] %
  is computed, and one of the roots, say $λ_i$, is chosen arbitrarily
  as \emph{positive direction}.
\end{enumerate}

The condition on the $ℓ_i$'s guarantees that each graph of
$ℓ_i$-isogenies on $\Ell_q(\O)$ is $2$-regular. %
The choice of a \emph{positive direction} allows us to \emph{orient}
the graph, by associating to $λ_i$ the isogeny with kernel
$E[ℓ_i]∩\ker(π-λ_i)$. %
The key exchange now proceeds like the ordinary Diffie-Hellman
protocol:
\begin{enumerate}
\item Alice chooses a random walk made of steps in $L$ along the
  positive direction; denote the walk by $ρ_A∈L^*$, and denote by
  $E_A=ρ_A(E)$ the curve where the walk terminates. %
  Note that $E_A$ only depends on how many times each $ℓ_i$ appears in
  $ρ_A$, and not on their order.
\item Bob does the same, choosing a random walk $ρ_B$ and computing
  $E_B=ρ_B(E)$.
\item Alice and Bob exchange $E_A$ and $E_B$.
\item Alice computes the \emph{shared secret} $ρ_A(E_B)$.
\item Bob computes the \emph{shared secret} $ρ_B(E_A)$.
\end{enumerate}
The actual computations are carried out by solving \emph{explicit
  isogeny problems} (see Problem~\ref{prob:explicit-isog}), in much
the same way they are done in the Elkies case of the SEA algorithm
(see Section~\ref{sec:appl-point-count}). %
The protocol is summarized in Figure~\ref{fig:rns}.

\begin{figure}
  \centering
  \begin{tabular}{l *{2}{p{30ex}<{\centering}}}
    \hline
    Public parameters & \multicolumn{2}{l}{An elliptic curve $E$ over a finite field $\F_q$,}\\
                      & \multicolumn{2}{l}{$D_π$, the discriminant of the Frobenius endomorphism of $E$,}\\
                      & \multicolumn{2}{l}{A set of primes $L=\{ℓ_1,\dots,ℓ_m\}$ such that $\left(\frac{D_π}{ℓ_i}\right)=1$,}\\
                      & \multicolumn{2}{l}{A \emph{Frobenius eigenvalue} $λ_i$ for each $ℓ_i$,}\\
    \hline
                      & {\bf Alice} & {\bf Bob}\\
    \hline
    Pick random secret & $ρ_A∈L^*$ & $ρ_B∈L^*$\\
    Compute public data & $E_A = ρ_A(E)$ & $E_B = ρ_B(E)$\\
    Exchange data &  \hfill $E_A \longrightarrow$ & $\longleftarrow E_B$ \hfill\strut \\
    Compute shared secret & $E_{AB} = ρ_A(E_B)$ & $E_{AB} = ρ_B(E_A)$
  \end{tabular}
  
  \caption{Rostovtsev-Stolbunov key exchange protocol based on random
    walks in an isogeny graph.}
  \label{fig:rns}
\end{figure}

We conclude this section with a discussion on the security of the
Rostovtsev-Stolbunov protocol. %
All the protocol's security rests on the isogeny path problem: given
$E$ and $E_A$, find an isogeny $ϕ:E\to E_A$ of smooth order. %
To be safe against exhaustive search and meet in the middle attacks as
seen in Section~\ref{sec:isog-graphs-crypt}, the set $\Ell_q(\O)$ must
be large. %
On average $\#\Ell_q(\O)\sim \sqrt{q}$, thus we shall take
$\log_2 q≈512$ for a security level of at most $128$ bits. %
However, some isogeny classes are much smaller than average, this is
why we also need check that $D_π$ has a large prime factor.

Furthermore, for the public and private curves to be (almost)
uniformly distributed in $\Ell_q(\O)$, we need the isogeny graph to be
connected; equivalently, we need the ideals $(ℓ_i,π-λ_i)$ to generate
$\Cl(\O)$. %
Theorem~\ref{th:ord-exp} ensures this is the case if
$\#L\sim (\log q)^2$, however it is usually sufficient to take a much
smaller set in practice. %
It is not enough to have an expander: we also need the random walks to
be longer than the mixing length, that is $\sim\log q$. %
And, since the key space grows exponentially with $\#L$, rather than
with the walk length, we shall also ask that
$\#L\sim \log q/\loglog q$.

When all conditions are met, the best known attack against this
cryptosystem is the meet in the middle strategy, which runs in
$O(\sqrt[4]{q})$ steps. %
However, the real case for this system is made by looking at attacks
performed on a \emph{quantum computer}. %
It is well known that Shor's algorithm~\cite{shor1994algorithms}
breaks the Diffie-Hellman cryptosystem in polynomial time on a quantum
computer, and thus it also breaks the protocol of
Figure~\ref{fig:walk-dh}. %
More generally, Shor's algorithm can solve the (generalized) discrete
logarithm problem in any Abelian group, and in particular in
$\Cl(\O)$. %
However, in the Rostovtsev-Stolbunov protocol, the attacker only sees
$E$, $E_A$ and $E_B$. %
Since there is no canonical way to map the curves to elements of
$\Cl(\O)$, it is not enough to be able to solve discrete logarithms in
it.

Childs, Jao and Soukharev~\cite{childs2014constructing} have shown how
to adapt quantum algorithms by Regev~\cite{regev04} and
Kuperberg~\cite{Kup} to solve the ordinary isogeny path problem in
subexponential time. %
Although their attack does not qualify as a total break, it makes the
Rostovtsev-Stolbunov protocol even less practical. %
Indeed, the protocol is already very slow, mainly due to the
relatively large size of the isogeny degree set $L$. %
If parameter sizes must be further enlarged to protect against quantum
attacks, it seems plausible that the Rostovtsev-Stolbunov protocol may
never be used in practice.

\subsection{Supersingular Isogeny Diffie-Hellman}

We finally come to the last cryptographic construction from isogeny
graphs. %
Compared to the ordinary case, graphs of supersingular isogenies have
two attractive features for constructing key exchange protocols. %
First, one isogeny degree is sufficient to obtain an expander graph;
by choosing one small prime degree, we have the opportunity to
construct more efficient protocols. %
Second, there is no action of an Abelian group, such as $\Cl(\O)$, on
them; it thus seems harder to use quantum computers to speed up the
supersingular isogeny path problem.

But the absence of a group action also makes it impossible to directly
generalize the Rostovtsev-Stolbunov protocol to supersingular
graphs. %
It turns out, however, that there is an algebraic structure acting on
supersingular graphs. %
We have seen that, if $E$ is a supersingular curve defined over $\F_p$
or $\F_{p^2}$, its endomorphism ring is isomorphic to an order in the
quaternion algebra $ℚ_{p,∞}$ ramified at $p$ and at infinity. %
There is more: supersingular curves are in correspondence with the
maximal orders of $ℚ_{p,∞}$, and their left ideals act on the graph
like isogenies. %
It would be rather technical to go into the details of the theory of
quaternion algebras and their maximal orders; instead, we describe the
key exchange protocol using only the language of isogenies, with the
\emph{caveat} that its security can only be properly evaluated by
also looking at its quaternion counterpart. %
The interested reader will find more details on quaternion algebras
in~\cite{waterhouse69,pizer1,pizer2,kohel,belding08-thesis,kohel2014quaternion}.

\begin{figure}
  \centering
  \begin{tikzpicture}
    \def\graph{
      \begin{scope}[every node/.style={fill,black,circle,inner sep=2pt}]
        \node at (0,0)  (1){};
        \node at (0,4) (20){};
        \node at (2,1)  (16z){};
        \node at (-2,1)  (81z){};
        \node at (-1,2) (77z){};
        \node at (1,2)  (20z){};
        \node at (-2,3)  (85z){};
        \node at (2,3)  (12z){};
      \end{scope}
    }
    
    \graph
    \begin{scope}[blue,every loop/.style={looseness=50}]
      \path (1) edge (20) edge (16z) edge (81z);
      \path (20) edge[loop left] (20) edge[loop right] (20);
      \path (16z) edge (81z) edge (77z);
      \path (81z) edge (20z);
      \path (77z) edge (20z) edge (85z);
      \path (20z) edge (12z);
      \path (12z) edge[bend right=10] (85z) edge[bend left=10] (85z);
    \end{scope}
        
    \begin{scope}[xshift=6cm]
      \graph
      \begin{scope}[red]
        \path (1) edge (85z) edge (81z) edge (12z) edge (16z);
        \path (20) edge (85z) edge (77z) edge (20z) edge (12z);
        \path (81z) edge (85z) edge (77z) edge (16z);
        \path (85z) edge (12z);
        \path (12z) edge (16z);
        \path (16z) edge (20z);
        \path (20z) edge[bend right=10] (77z) edge[bend left=10] (77z);
      \end{scope}
    \end{scope}
  \end{tikzpicture}
  \caption{Supersingular isogeny graphs of degree 2 (left, blue) and 3
    (right, red) on $\F_{97^2}$.}
  \label{fig:sup-97-2-3}
\end{figure}
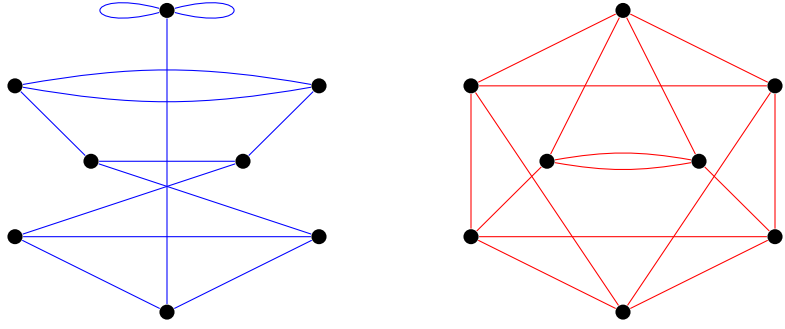

The key idea of the Supersingular Isogeny Diffie-Hellman protocol
(SIDH), first proposed in~\cite{jao+defeo2011}, is to let Alice and
Bob take random walks in two distinct isogeny graphs on the same
vertex set. %
In practice, we choose a \emph{large enough} prime $p$, and two
\emph{small} primes $ℓ_A$ and $ℓ_B$. %
The vertex set is going to consist of the supersingular $j$-invariants
defined over $\F_{p^2}$, Alice's graph is going to be made of
$ℓ_A$-isogenies, while Bob is going to use $ℓ_B$-isogenies. %
Figure~\ref{fig:sup-97-2-3} shows a toy example of such graphs, where
$p=97$, $ℓ_A=2$ and $ℓ_B=3$.

Even this, though, is not sufficient to define a key exchange
protocol, because there is no canonical way of labeling the edges of
these graphs. %
We shall introduce, then, a very \emph{ad hoc} construction leveraging
the group structure of elliptic curves. %
Recall that a separable isogeny is uniquely defined by its kernel, and
that in this case $\deg ϕ = \#\ker ϕ$. %
More precisely, a walk of length $e_A$ in the $ℓ_A$-isogeny graph
corresponds to a kernel of size $ℓ_A^{e_A}$; and this kernel is cyclic
if and only if the walk \emph{does not backtrack}. %

Hence, Alice choosing a secret walk of length $e_A$ is equivalent to
her choosing a secret cyclic subgroup $〈A〉⊂E[ℓ_A^{e_A}]$. %
If we let Alice choose one such subgroup, and Bob choose similarly a
secret $〈B〉⊂E[ℓ_B^{e_B}]$, then there is a well defined subgroup
$〈A〉+〈B〉=〈A,B〉$, defining an isogeny to $E/〈A,B〉$. %
Since we have taken care to choose $ℓ_A≠ℓ_B$, the group $〈A,B〉$ is
cyclic of order $ℓ_A^{e_A}ℓ_B^{e_B}$. %
This is illustrated in Figure~\ref{fig:sidh-diag}.

\begin{figure}
  \centering
  \begin{tikzpicture}
    \begin{scope}
      \draw (0,1.2) node[anchor=east] {$\bl{\ker α=〈A〉}⊂ E[\ell_A^{e_A}]$};
      \draw (0,0.4) node[anchor=east] {$\rd{\ker β=〈B〉}⊂ E[\ell_B^{e_B}]$};
      \draw (0,-0.4) node[anchor=east] {$\ker α' = 〈\rd{β}\bl{(A)}〉$};
      \draw (0,-1.2) node[anchor=east] {$\ker β' = 〈\bl{α}\rd{(B)}〉$};
    \end{scope}
    \begin{scope}[xshift=4.5cm]
      \large
      \node[matrix of nodes, ampersand replacement=\&, column sep=3cm, row sep=1.5cm] (diagram) {
        |(E)| $E$ \& |(Es)| $E/〈\bl{A}〉$ \\
        |(Ep)| {$E/〈\rd{B}〉$} \& |(Eps)| {$E/〈\bl{A},\rd{B}〉$}\\
      };
      \path[->,blue] (E) edge node[auto] {$α$} (Es);
      \path[->] (Ep) edge node[auto,swap] {$α'$} (Eps);
      \path[->,red] (E) edge node[auto,swap] {$β$} (Ep);
      \path[->] (Es) edge node[auto] {$β'$} (Eps);
    \end{scope}
  \end{tikzpicture}
  \caption{Commutative isogeny diagram constructed from Alice's and
    Bob's secrets. %
    Quantities known to Alice are drawn in blue, those known to Bob
    are drawn in red.}
  \label{fig:sidh-diag}
\end{figure}

At this point, we would like to define a protocol where Alice and Bob
choose random cyclic subgroups $〈A〉$ and $〈B〉$ in some \emph{large
  enough} torsion groups, and exchange enough information to both
compute $E/〈A,B〉$ (up to isomorphism), without revealing their
respective secrets. %
We are faced with two difficulties, though:
\begin{enumerate}
\item The points of $〈A〉$ (or $〈B〉$) may not be rational. %
  Indeed, in general they may be defined over a field extension of
  degree as large as $ℓ_A^{e_A}$, thus requiring an exponential amount
  of information to be explicitly represented.
\item The diagram in Figure~\ref{fig:sidh-diag} shows no way by which
  Alice and Bob could compute $E/〈A,B〉$ without revealing their
  secrets to each other.
\end{enumerate}

We will solve both problems by carefully controlling the group
structure of our supersingular curves. %
This is something that is very hard to do in the ordinary case, but
totally elementary in the supersingular one, as the following
proposition shows.

\begin{theorem}[Group structure of supersingular curves]
  Let $p$ be a prime, and let $E$ be a supersingular curve defined
  over a finite field $\F_q$ with $q=p^m$ elements. %
  Let $t$ be the trace of the Frobenius endomorphism of $E/k$, then
  one of the following is true:
  \begin{itemize}
  \item $m$ is odd and
    \begin{itemize}
    \item $t=0$, or
    \item $p=2$ and $t^2=2q$, or
    \item $p=3$ and $t^2=3q$;
    \end{itemize}
  \item $m$ is even and
    \begin{itemize}
    \item $t^2=4q$, or
    \item $t^2=q$, and $j(E)=0$, and $E$ is not isomorphic to
      $y^2=x^3±1$, or
    \item $t^2=0$, and $j(E)=1728$, and $E$ is not isomorphic to
      $y^2=x^3±x$.
    \end{itemize}
  \end{itemize}
  
  The group structure of $E(\F_q)$ is one of the following:
  \begin{itemize}
  \item If $t^2=q,2q,3q$, then $E(\F_q)$ is cyclic;
  \item If $t=0$, then $E(\F_q)$ is either cyclic, or isomorphic to
    $ℤ/\frac{q+1}{2}ℤ⊕ℤ/2ℤ$;
  \item If $t=∓2\sqrt{q}$, then $E(\F_q)≃(ℤ/(\sqrt{q}±1)ℤ)^2$.
  \end{itemize}
\end{theorem}
\begin{proof}
  See~\cite{waterhouse69,MOV}.
\end{proof}

Of all the cases, the only one we are concerned with is $q=p^2$, and
$E(\F_q)≃(ℤ/(p±1)ℤ)^2$. %
Since we have full control on $p$, we can choose it so that $E(\F_q)$
contains two large subgroups $E[ℓ_A^{e_A}]$ and $E[ℓ_B^{e_B}]$ of
coprime order. %
Hence, once $ℓ_A^{e_A}$ and $ℓ_B^{e_B}$ are fixed, we look for a prime
of the form $p=ℓ_A^{e_A}ℓ_B^{e_B}f∓1$, where $f$ is a small
cofactor. %
In practice, such primes are abundant, and we can easily take $f=1$. %
This solves the first problem: $E(\F_q)$ now contains
$ℓ_A^{e_A-1}(ℓ_A+1)$ cyclic subgroups of order $ℓ_A^{e_A}$, each
defining a distinct isogeny; hence, a single point $A∈E(\F_q)$ is
enough to represent an isogeny walk of length $e_A$.

The second problem is solved by a very peculiar trick, which sets SIDH
apart from other isogeny based protocols. %
The idea is to let Alice and Bob publish some additional information
to help each other compute the shared secret. %
Let us summarize what are the quantities known to Alice and Bob. %
To set up the cryptosystem, they have publicly agreed on a prime $p$
and a supersingular curve $E$ such that
\[E(\F_{p^2}) ≃ (ℤ/ℓ_A^{e_A}ℤ)^2⊕(ℤ/ℓ_B^{e_B}ℤ)^2⊕(ℤ/fℤ)^2.\] %
It will be convenient to also fix public bases of their respective
torsion groups:
\begin{align*}
  E[ℓ_A^{e_A}] &= 〈P_A,Q_A〉,\\
  E[ℓ_B^{e_B}] &= 〈P_B,Q_B〉.
\end{align*}
To start the protocol, they choose random secret subgroups
\begin{align*}
  〈A〉 &= 〈[m_A]P_A+[n_A]Q_A〉 ⊂ E[ℓ_A^{e_A}],\\  
  〈B〉 &= 〈[m_B]P_B+[n_B]Q_B〉 ⊂ E[ℓ_B^{e_B}],
\end{align*}
of respective orders $ℓ_A^{e_A},ℓ_B^{e_B}$, and compute the secret
isogenies
\begin{align*}
  α : E &\to E/〈Α〉,\\
  β : E &\to E/〈B〉.
\end{align*}
They respectively publish $E_A=E/〈Α〉$ and $E_B=E/〈B〉$. %

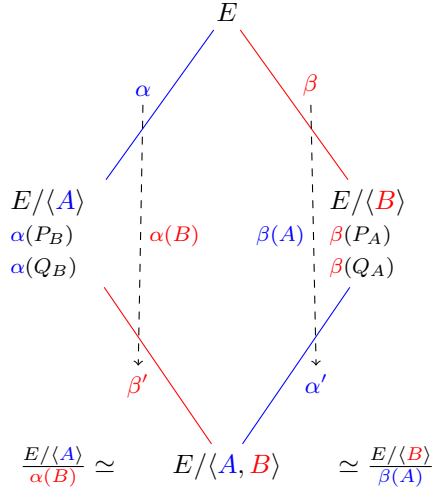
\begin{figure}
  \centering
  \begin{tikzpicture}
    \node[matrix of nodes, ampersand replacement=\&, column sep=4mm, row sep=2cm] (diagram) {
      \& |(1)| $E$ \\
      |(a)| \parbox{1.5cm}{$E/〈\bl{A}〉$\\{\footnotesize $\bl{α}(P_B)\\\bl{α}(Q_B)$}} \& \&
      |(b)| \parbox{1.5cm}{$E/〈\rd{B}〉$\\{\footnotesize $\rd{β}(P_A)\\\rd{β}(Q_A)$}}\\
      \normalsize $\frac{E/〈\bl{A}〉}{\rd{α(B)}} \simeq$ \&
      |(ab)|  $E/〈\bl{A},\rd{B}〉$ \&
      \normalsize $\simeq \frac{E/〈\rd{B}〉}{\bl{β(A)}}$\\
    };
    \small
    \path[blue] (1) edge node[auto,swap](phia) {$α$} (a);
    \path[red] (1) edge node[auto](phib) {$β$} (b);
    \path[red] (a) edge node[auto,swap](psia){$β'$} (ab);
    \path[blue] (b) edge node[auto](psib){$α'$} (ab);
    \path[dashed,->] (phia) edge node[auto]{\footnotesize $\rd{α(B)}$} (psia);
    \path[dashed,->] (phib) edge node[auto,swap]{\footnotesize $\bl{β(A)}$} (psib);
  \end{tikzpicture}

  \caption{Schematics of SIDH key exchange. Quantities only known to
    Alice are drawn in blue, quantities only known to Bob in red.}
  \label{fig:sidh}
\end{figure}

Now, to compute the shared secret $E/〈A,B〉$, Alice needs to compute
the isogeny $α':E/〈B〉\to E/〈A,B〉$, which kernel is generated by
$β(A)$. %
We see that the kernel of $α'$ depends on both secrets, thus Alice
cannot compute it without Bob's assistance. %
The trick here is for Bob to publish the values $β(P_A)$ and $β(Q_A)$:
they do not require the knowledge of Alice's secret, and it is
conjectured that they do not give any advantage in computing
$E/〈A,B〉$ to an attacker. %
From Bob's published values, Alice can compute $β(A)$ as
$[m_A]β(P_A) + [n_A]β(Q_A)$, and complete the protocol. %
Bob performs the analogous computation, with the help of Alice. %
The protocol is summarized in Figure~\ref{fig:sidh-prot}, and
schematized in Figure~\ref{fig:sidh}.

\begin{figure}
  \centering
  \begin{tabular}{l *{2}{p{32ex}<{\centering}}}
    \hline
    Public parameters & \multicolumn{2}{l}{Primes $ℓ_A,ℓ_B$, and a prime $p=ℓ_A^{e_A}ℓ_B^{e_B}f∓1$,}\\
                      & \multicolumn{2}{l}{A supersingular elliptic curve $E$ over $\F_{p^2}$ of order $(p±1)^2$,}\\
                      & \multicolumn{2}{l}{A basis $〈P_A,Q_A〉$ of $E[ℓ_A^{e_A}]$,}\\
                      & \multicolumn{2}{l}{A basis $〈P_B,Q_B〉$ of $E[ℓ_B^{e_B}]$,}\\
    \hline
                      & {\bf Alice} & {\bf Bob}\\
    \hline
    Pick random secret & $A=[m_A]P_A+[n_A]Q_A$ & $B=[m_B]P_B+[n_B]Q_B$\\[1ex]
    Compute secret isogeny & $α:E\to E_A=E/〈A〉$ & $β:E\to E_B=E/〈B〉$\\[1ex]
    Exchange data &  \hfill $E_A,α(P_B),α(Q_B) \longrightarrow$ & $\longleftarrow E_B,β(P_A),β(Q_A)$ \hfill\strut \\[1ex]
    Compute shared secret & $E/〈A,B〉 = E_B/〈β(A)〉$ & $E/〈A,B〉 = E_A/〈α(B)〉$
  \end{tabular}
  
  \caption{Supersingular Isogeny Diffie-Hellman key exchange protocol.}
  \label{fig:sidh-prot}
\end{figure}

We end with a discussion on parameter sizes. %
It is clear that the key space of SIDH depends on the size of the
subgroups $E[ℓ_A^{e_A}]$ and $E[ℓ_B^{e_B}]$, hence we must take
$ℓ_A^{e_A}\sim ℓ_B^{e_B}$ so to make attacks equally hard against
Alice or Bob's public data. %
However this puts serious constraints on the isogeny walks performed
in SIDH. %
Indeed, we have seen that the size of the supersingular isogeny graph
is $O(p)$, whereas the size of Alice's (or Bob's) public key space is
only $O(\sqrt{p})$. %
Said otherwise, Alice and Bob take random walks \emph{much shorter}
than the diameter of the graph. %
At the moment, it is not clear how this affects the security of the
protocol.

To choose an appropriate size for $p$, we start by looking at attacks
that only use the $j$-invariants published by Alice and Bob. %
Given curves $E$ and $E_A$, connected by an isogeny of degree
$ℓ_A^{e_A}$, an easy variation on the meet-in-the-middle paradigm
finds the secret isogeny in $O(ℓ_A^{e_A/2})$ steps (and
$O(ℓ_A^{e_A/2})$ storage) as follows: tabulate all possible walks of
length $\lfloor e_A/2\rfloor$ starting from $E$, then iterate over the
walks of length $\lceil e_A/2\rceil$ starting from $E_A$, until a
collision is found. %
The same collision can also be found with $O(ℓ_A^{e_A/3})$ queries to
a quantum oracle, using a quantum algorithm due to
Tani~\cite{tani2009claw}. %
Because the isogeny walks are shorter than the diameter, we expect to
find only one collision, and that is precisely Alice's secret isogeny.

It turns out these are the best known attacks against SIDH, even
taking into account the additional information passed by Alice and
Bob. %
Hence, taking $\log_2p=n$ offers a classical security of $\sim n/4$
bits, and a quantum security of $\sim n/6$ qubits. %
In conclusion, to obtain a 128-qubit and 192-bit secure system, we
would have to find a 768-bit prime of the for
$p=ℓ_A^{e_A}ℓ_B^{e_B}f±1$, with $e_A\log_2ℓ_A\sim e_B\log_2ℓ_B\sim384$. %
In practice, we usually take $ℓ_A=2$ and $ℓ_B=3$ for efficiency
reasons, and an example of one such prime is $p=2^{387}3^{242}-1$.

\section{Further topics in isogeny based cryptography}

We conclude these notes with a brief overview of the current research
topics in isogeny based cryptography. %
We only focus on constructions derived from supersingular isogenies,
as they currently are the most promising ones.

\paragraph{Efficient implementation of SIDH}
What makes SIDH interesting is its relatively good efficiency,
especially when compared with other isogeny based protocols. %
However, several optimizations are required in order to achieve a
compact and fast implementation, competitive with other post-quantum
key-exchange candidates. %
In short, one must optimize each of these levels:
\begin{itemize}
\item The arithmetic of $\F_p$ benefits from the special form of
  $p$, especially for primes of the form $p=2^a3^b-1$, as explained
  in~\cite{costello2016sidh,vercauteren-sidh-fp,cryptoeprint:2016:986};
\item The arithmetic of $\F_{p^2}$ benefits from the fact that $-1$ is
  not a square in $\F_p$, whenever $p=-1\mod 2$;
\item The arithmetic of elliptic curves benefits from using Montgomery
  models, and optimized formulas for doublings, triplings, scalar
  multiplications and
  isogenies~\cite{defeo+jao+plut12,costello2016sidh,cryptoeprint:2017:504,cryptoeprint:2017:1015};
\item Field inversions can be avoided using projective coordinates and
  \emph{projectivized curve equations}~\cite{costello2016sidh};
\item The full computation and evaluation of the secret isogeny from a
  generator of its kernel must be performed using a quasi-linear
  algorithm first described in~\cite{defeo+jao+plut12}.
\end{itemize}

Undoubtedly, the latter is the most novel and surprising of the
optimizations. %
For lack of space, we do not describe any of them here, and we
primarily address the interested reader to~\cite{defeo+jao+plut12}
and~\cite{costello2016sidh}.

By putting together all the optimizations mentioned above, the SIDH
scheme can be made relatively practical, as shown
in~\cite{defeo+jao+plut12,costello2016sidh}, although one or two
orders of magnitude slower than other post-quantum competitors. %
Where SIDH really excels, is in its \emph{very short key sizes},
actually the shortest among post-quantum candidates, at the time of
writing. %
This key size can be shrunk even more through \emph{key compression}
techniques~\cite{azarderakhsh2016key,Costello2017}. %
However, the size of the isogeny graph in SIDH is much larger than the
size of the key space, it is thus, in principle, possible to make even
shorter keys; how to do this efficiently is still an open question.

\paragraph{Security of SIDH}
We can formally state the security of SIDH as a hardness assumption on
a problem called SSDDH. %
As mentioned previously, the best known algorithms for SSDDH have
exponential complexity, even on a quantum computer.

\begin{problem}[Supersingular Decision Diffie-Hellman]
  Let $E,ℓ_A,ℓ_B,e_A,e_B,P_A,Q_A,P_B,Q_B$ be the parameters of an SIDH
  protocol.

  Given a tuple sampled with probability $1/2$ from one of
  the following two distributions:
  \begin{enumerate} 
  \item $(E/〈A〉, ϕ(P_B), ϕ(Q_B), E/〈B〉, ψ(P_A), ψ(Q_A), E/〈A,B〉)$, where
    \begin{itemize}
    \item $A∈E$ is a uniformly random point of order $ℓ_A^{e_A}$,
    \item $B∈E$ is a uniformly random point of order $ℓ_B^{e_B}$,
    \item $ϕ:E\to E/〈A〉$ is the isogeny of kernel $〈A〉$, and
    \item $ψ:E \to E/〈B〉$ is the isogeny of kernel $〈B〉$;
    \end{itemize}
  \item $(E/〈A〉, ϕ(P_B), ϕ(Q_B), E/〈B〉, ψ(P_A), ψ(Q_A), E/〈C〉)$,
    where $A,B,ϕ,ψ$ are as above, and where $C∈E$ is a uniformly random
    point of order $ℓ_A^{e_A}ℓ_B^{e_B}$;
  \end{enumerate}
  determine from which distribution the tuple is sampled.
\end{problem}

Assuming SSDDH is hard, we can formally prove the security of the key
exchange against \emph{passive adversaries}, i.e., those adversaries
who can see all messages sent between Alice and Bob, but who do not
modify them. %
We address the interested reader to~\cite{defeo+jao+plut12} for the
technical details.

It is apparent that SSDDH is a very special instance of the isogeny
path problem; it is thus conceivable that specially crafted algorithms
could break SIDH without solving the generic isogeny path problem. %
As an illustration, consider the following problem.

\begin{problem}
  Let $E,ℓ_A,ℓ_B,e_A,e_B,P_A,Q_A,P_B,Q_B$ be the parameters of an SIDH
  protocol.

  Let $A∈E$ be a point of order $ℓ_A^{e_A}$, and let $ϕ:E\to E/〈A〉$. %
  Given $E/〈Α〉$, $ϕ(P_B)$ and $ϕ(Q_B)$ compute $ϕ(R)$ for an
  arbitrary point $R∈E$ of order $ℓ_A^{e_A}$.
\end{problem}

It is easy to verify that solving this problem immediately reveals the
secret $〈A〉$. Indeed, $ϕ(R)$ is an element of $\ker\hat{ϕ}$, from
which we can recover $\hat{ϕ}$ and $ϕ$, and thus $〈A〉$. %
An efficient solution to this problem completely breaks SIDH, without
doing anything for the generic isogeny path problem.%
\footnote{The converse reduction is not evident either: given an
  oracle solving the isogeny path problem, how can we break SIDH? %
  A partial answer is given
  in~\cite{kohel2014quaternion,galbraithsecurity}, where it is shown
  that, knowing the endomorphism rings of $E$ and $E/〈A〉$, an
  attacker can solve the isogeny path problem, and then break SIDH, in
  polynomial time.}

And indeed, although the security of SSDDH is still unblemished at the
time of writing, several polynomial-time attacks have appeared against
variations of SIDH. %
The interested reader will find more details in the following
references:
\begin{itemize}
\item A key-recovery attack against a \emph{static key} version of
  SIDH, where Alice uses a long term secret
  isogeny~\cite{galbraithsecurity};
\item Key-recovery attacks in various \emph{leakage
    models}~\cite{galbraithsecurity,gelin2017loop,ti2017fault};
\item Key recovery attacks against some \emph{unbalanced} variants
  of SIDH~\cite{cryptoeprint:2017:571}.
\end{itemize}
Finally, it is worth mentioning that there is a quantum subexponential
attack~\cite{biasse2014quantum} in the case where both $E$ and
$E/〈A〉$ are defined over $\F_p$.

\paragraph{Other protocols}
Key exchange is not the only public-key protocol that can be derived
from isogeny graphs. %
It is easy, for example, to derive a public-key encryption protocol
similar to El Gamal from either the Rostovtsev-Stolbunov protocol or
SIDH. %
We illustrate the second:
\begin{itemize}
\item Alice's secret key is an isogeny $α:E\to E/〈A〉$; her public
  key contains $E/〈A〉$ and the evaluation of $α$ on \emph{Bob's
    basis} $〈P_B,Q_B〉$.
\item To encrypt a message $m$, Bob chooses a random $β:E\to E/〈B〉$,
  and computes the shared secret $E/〈A,B〉$, which he converts to a
  binary string $s$ (e.g., by hashing the $j$ invariant of
  $E/〈A,B〉$); he sends to Alice the message
  $(E/〈B〉,β(P_A),β(Q_A),m⊕s)$.
\item To decrypt, Alice uses $E/〈B〉,β(P_A),β(Q_A)$ to compute the
  shared secret $E/〈A,B〉$, which she converts to $s$, and finally
  she unmasks $m⊕s$.
\end{itemize}

In~\cite{defeo+jao+plut12}, it is proven that this protocol is IND-CPA
secure under the SSDDH assumption. %
Achieving IND-CCA security is harder, as the attack against static
keys in~\cite{galbraithsecurity} shows, however it is possible to
apply a generic transformation to obtain an IND-CCA secure \emph{key
  encapsulation mechanism}.

One may expect that digital signatures would also generalize easily to
the isogeny setting, but both Schnorr signatures and ECDSA rely on the
existence of a group law on the public data, something that is missing
both in the ordinary and in the supersingular case.

\begin{figure}
  \centering
  \begin{tabular}{l *{2}{p{32ex}<{\centering}}}
    \hline
    Parameters & \multicolumn{2}{l}{Primes $ℓ_A,ℓ_B$, and a prime $p=ℓ_A^{e_A}ℓ_B^{e_B}f∓1$,}\\
               & \multicolumn{2}{l}{A supersingular elliptic curve $E$ over $\F_{p^2}$ of order $(p±1)^2$,}\\
               & \multicolumn{2}{l}{A basis $〈P_B,Q_B〉$ of $E[ℓ_B^{e_B}]$.}\\
    \hline
    Secret key & \multicolumn{2}{l}{An isogeny $α:E\to E/〈A〉$ of degree $ℓ_A^{e_A}$.}\\
    \hline
    Public key & \multicolumn{2}{l}{The curve $E/〈A〉$, the images $α(P_B),α(Q_B)$.}\\
    \hline
                      & {\bf Alice} & {\bf Bob}\\
    \hline
    Pick random & $B∈E[ℓ_B^{e_B}]$ of order $ℓ_B^{e_B}$ \\[1ex]
    Compute masking isogeny & $β:E\to E/〈B〉$\\[1ex]
    Commit &  \hfill $(E/〈B〉,E/〈A,B〉) \longrightarrow$\\[1ex]
    Challenge && $\longleftarrow b∈\{0,1\}$ \hfill\strut \\[1ex]
    Reveal & \hfill if $b=0$, send $(B,α(B)) \phantom{\longrightarrow}$\\[1ex]
               & \hfill if $b=1$, send $β(A) \longrightarrow$
  \end{tabular}
  
  \caption{Supersingular Isogeny Zero-Knowledge Identification protocol.}
  \label{fig:sidh-zk}
\end{figure}

To our rescue, comes a zero-knowledge protocol based on the same
construction shown in Figure~\ref{fig:sidh-diag}. %
In this protocol, Alice's secret key is an isogeny $α:E\to E/〈A〉$;
her public key is the curve $E/〈A〉$, together with a description of
the action of $α$ on $E[ℓ_B^{e_B}]$, as in SIDH. %
To prove knowledge of $α$ to Bob, she takes a random subgroup
$〈B〉⊂E[ℓ_B^{e_B}]$, computes a commutative diagram as in
Figure~\ref{fig:sidh-diag}, and sends to Bob the curves $E/〈B〉$ and
$E/〈A,B〉$. %
To verify that Alice knows the secret, Bob asks her one of two
questions at random:
\begin{itemize}
\item either reveal the point $B$ and its image $α(B)$,
\item or reveal the point $β(A)$.
\end{itemize}
After receiving Alice's answer, he accepts only if the points do
define isogenies between the curves $E,E/〈A〉,E/〈B〉,E/〈A,B〉$ as
expected. %
The protocol is summarized in Figure~\ref{fig:sidh-zk}.

Intuitively, if Alice respects the protocol, she always succeeds in
convincing Bob. %
If she cheats, she only has one chance out of two of guessing Bob's
challenge and succeed in tricking him. %
Thus, by iterating the protocol a sufficient number of times, a
cheater's chance of success can be made arbitrarily small at
exponential pace. %
The protocol is zero-knowledge because revealing $B$ and $α(B)$ does
not reveal anything that Bob does not already know. %
Revealing $β(A)$ is trickier, and we need to make one more security
assumption, named Decisional Supersingular Product (DSSP), to prove
zero knowledge. %
In~\cite{defeo+jao+plut12} it is proven that this protocol is secure
and zero-knowledge under the SSDDH\footnote{Actually, a weaker
  assumption named CSSI.} and DSSP assumptions.%
\footnote{The paper~\cite{defeo+jao+plut12} also hints at a variant of
  the zero-knowledge protocol where Bob challenges Alice to open one
  out of three commitments, namely one of $Β,α(B),β(A)$. %
  This variant is less efficient, since a cheater has $2/3$ chances of
  success, however its security relies on the stronger isogeny walk
  problem, rather than on SSDDH.}

Using a generic construction, such as the Fiat-Shamir
heuristic~\cite{fiat1986prove}, it is possible to derive a signature
scheme from the zero-knowledge protocol above. %
Alternative signature schemes based on the same construction, with
different desirable properties, are presented
in~\cite{cryptoeprint:2016:1154,cryptoeprint:2017:186}. %
However, all these protocols suffer from the high cost of having to
iterate hundreds of times the basic building block of
Figure~\ref{fig:sidh-zk}. %
Obtaining an efficient signature scheme from isogeny assumptions is
still an open problem.

More protocols can be obtained by slightly generalizing the SIDH
construction. %
If we allow the prime to be of the form
$p=ℓ_A^{e_A}ℓ_B^{e_B}ℓ_C^{e_C}±1$, we can construct a commutative cube
in the same way the square of Figure~\ref{fig:sidh-diag} was
constructed. %
Using primes of this form, Sun, Tian and Wang have proposed a
\emph{strong designated verifier} signature
scheme~\cite{sun2012toward}. %
Adding one more prime $ℓ_D$ in the mix, Jao and Soukharev have
proposed \emph{undeniable} signatures~\cite{jao2014isogeny}.
The drawback of all these schemes is that, as we add more torsion
subgroups to the base curve, the size of the primes grows, making the
schemes less and less practical. %

In general, isogeny graphs are much less flexible than the classical
discrete logarithm problem. %
Many of the protocols that have been built on discrete logarithms fail
to be ported to isogeny based cryptography. %
Devising new post-quantum protocols, retaining some of the desirable
properties of classical ones, is a very active area of research in
isogeny based cryptography.

\section*{Exercices}

\begin{exercice}
  Prove Proposition~\ref{th:graph-eigen}.
\end{exercice}

\begin{exercice}
  \label{ex:schreier}
  Show that a Schreier graph $(S⊂G, X)$ is an $ε$-expander if and only
  if $S$ generates $G$.
\end{exercice}

\begin{exercice}
  \label{ex:elgamal}
  Derive encryption protocols \emph{\`a la} El Gamal from the key
  exchange protocols of Section~\ref{sec:post-quantum-key}.
\end{exercice}

\clearpage
\bibliographystyle{plain}
\bibliography{refs}

\begin{thebibliography}{10}

\bibitem{atkin88}
Arthur O.~L. Atkin.
\newblock The number of points on an elliptic curve modulo a prime.
\newblock 1988.

\bibitem{atkin91}
Arthur O.~L. Atkin.
\newblock The number of points on an elliptic curve modulo a prime.
\newblock
  \url{http://www.lix.polytechnique.fr/Labo/Francois.Morain/AtkinEmails/19910614.txt},
  1991.

\bibitem{azarderakhsh2016key}
Reza Azarderakhsh, David Jao, Kassem Kalach, Brian Koziel, and Christopher
  Leonardi.
\newblock Key compression for isogeny-based cryptosystems.
\newblock In {\em Proceedings of the 3rd ACM International Workshop on ASIA
  Public-Key Cryptography}, pages 1--10. ACM, 2016.

\bibitem{belding08-thesis}
Juliana~V. Belding.
\newblock {\em Number Theoretic Algorithms for Elliptic Curves}.
\newblock PhD thesis, University of Maryland, 2008.

\bibitem{biasse2014quantum}
Jean-Fran{\c{c}}ois Biasse, David Jao, and Anirudh Sankar.
\newblock A quantum algorithm for computing isogenies between supersingular
  elliptic curves.
\newblock In {\em International Conference in Cryptology in India}, pages
  428--442. Springer, 2014.

\bibitem{cryptoeprint:2016:986}
Joppe~W. Bos and Simon Friedberger.
\newblock Fast arithmetic modulo $2^xp^y\pm 1$.
\newblock Cryptology ePrint Archive, Report 2016/986, 2016.
\newblock \url{http://eprint.iacr.org/2016/986}.

\bibitem{bostan+morain+salvy+schost08}
Alin Bostan, Fran\c{c}ois Morain, Bruno Salvy, and \'{E}ric Schost.
\newblock Fast algorithms for computing isogenies between elliptic curves.
\newblock {\em Math. Comp.}, 77:1755--1778, September 2008.

\bibitem{charles+lauter+goren09}
Denis~X. Charles, Eyal~Z. Goren, and Kristin~E. Lauter.
\newblock Cryptographic hash functions from expander graphs.
\newblock {\em Journal of Cryptology}, 22(1):93--113, January 2009.

\bibitem{childs2014constructing}
Andrew Childs, David Jao, and Vladimir Soukharev.
\newblock Constructing elliptic curve isogenies in quantum subexponential time.
\newblock {\em Journal of Mathematical Cryptology}, 8(1):1--29, 2014.

\bibitem{cryptoeprint:2017:504}
Craig Costello and Huseyin Hisil.
\newblock A simple and compact algorithm for {SIDH} with arbitrary degree
  isogenies.
\newblock Cryptology ePrint Archive, Report 2017/504, 2017.
\newblock \url{http://eprint.iacr.org/2017/504}.

\bibitem{Costello2017}
Craig Costello, David Jao, Patrick Longa, Michael Naehrig, Joost Renes, and
  David Urbanik.
\newblock {\em Efficient Compression of SIDH Public Keys}, pages 679--706.
\newblock Springer International Publishing, Cham, 2017.

\bibitem{costello2016sidh}
Craig Costello, Patrick Longa, and Michael Naehrig.
\newblock Efficient algorithms for {S}upersingular {I}sogeny
  {D}iffie-{H}ellman.
\newblock In Matthew Robshaw and Jonathan Katz, editors, {\em Advances in
  Cryptology -- CRYPTO 2016: 36th Annual International Cryptology Conference},
  pages 572--601. Springer Berlin Heidelberg, 2016.

\bibitem{couveignes94}
Jean-Marc Couveignes.
\newblock {\em {Quelques calculs en th{\'{e}}orie des nombres}}.
\newblock PhD thesis, Universit\'{e} de Bordeaux, 1994.

\bibitem{couveignes96}
Jean-Marc Couveignes.
\newblock Computing $\ell$-isogenies using the $p$-torsion.
\newblock In {\em ANTS-II: Proceedings of the Second International Symposium on
  Algorithmic Number Theory}, pages 59--65, London, UK, 1996. Springer-Verlag.

\bibitem{couveignes00}
Jean-Marc Couveignes.
\newblock Isomorphisms between {A}rtin-{S}chreier towers.
\newblock {\em Mathematics of Computation}, 69(232):1625--1631, 2000.

\bibitem{Couv}
Jean-Marc Couveignes.
\newblock Hard homogeneous spaces.
\newblock \url{http://eprint.iacr.org/2006/291/}, 2006.

\bibitem{couveignes+lercier11}
Jean-Marc Couveignes and Reynald Lercier.
\newblock Fast construction of irreducible polynomials over finite fields.
\newblock {\em Israel Journal of Mathematics}, 194(1):77--105, 2013.

\bibitem{cox2011primes}
David~A Cox.
\newblock {\em Primes of the form $x^2+ny^2$: Fermat, class field theory, and
  complex multiplication}, volume~34.
\newblock John Wiley \& Sons, 2011.

\bibitem{df+thesis}
Luca De~Feo.
\newblock {\em {A}lgorithmes {R}apides pour les {T}ours de {C}orps {F}inis et
  les {I}sog{\'{e}}nies}.
\newblock PhD thesis, Ecole Polytechnique X, December 2010.

\bibitem{DeDoSc13}
Luca De~Feo, Javad Doliskani, and {\'E}ric Schost.
\newblock Fast algorithms for $\ell$-adic towers over finite fields.
\newblock In {\em ISSAC'13: Proceedings of the 38th International Symposium on
  Symbolic and Algebraic Computation}, pages 165--172. ACM, 2013.

\bibitem{defeo2016explicit}
Luca De~Feo, Cyril Hugounenq, J{\'e}r{\^o}me Pl{\^u}t, and {\'E}ric Schost.
\newblock Explicit isogenies in quadratic time in any characteristic.
\newblock {\em LMS Journal of Computation and Mathematics}, 19(A):267--282,
  2016.

\bibitem{defeo+jao+plut12}
Luca De~Feo, David Jao, and J{\'e}r{\^o}me Pl{\^u}t.
\newblock Towards quantum-resistant cryptosystems from supersingular elliptic
  curve isogenies.
\newblock {\em Journal of Mathematical Cryptology}, 8(3):209--247, 2014.

\bibitem{df+schost09}
Luca De~Feo and \'{E}ric Schost.
\newblock Fast arithmetics in {A}rtin-{S}chreier towers over finite fields.
\newblock In {\em ISSAC '09: Proceedings of the 2009 international symposium on
  Symbolic and algebraic computation}, pages 127--134, New York, NY, USA, 2009.
  ACM.

\bibitem{dh}
Whitfield Diffie and Martin~E. Hellman.
\newblock New directions in cryptography.
\newblock {\em IEEE Transactions on Information Theory}, IT-22(6):644--654,
  1976.

\bibitem{elkies92}
Noam~D. Elkies.
\newblock Explicit isogenies.
\newblock 1992.

\bibitem{elkies98}
Noam~D. Elkies.
\newblock Elliptic and modular curves over finite fields and related
  computational issues.
\newblock In {\em Computational perspectives on number theory (Chicago, IL,
  1995)}, volume~7 of {\em Studies in Advanced Mathematics}, pages 21--76,
  Providence, RI, 1998. AMS International Press.

\bibitem{cryptoeprint:2017:1015}
Armando Faz-Hern\'andez, Julio L\'opez, Eduardo Ochoa-Jim\'enez, and Francisco
  Rodr\'iguez-Henr\'iquez.
\newblock A faster software implementation of the supersingular isogeny
  diffie-hellman key exchange protocol.
\newblock Cryptology ePrint Archive, Report 2017/1015, 2017.
\newblock \url{http://eprint.iacr.org/2017/1015}.

\bibitem{fiat1986prove}
Amos Fiat and Adi Shamir.
\newblock How to prove yourself: Practical solutions to identification and
  signature problems.
\newblock In {\em Conference on the Theory and Application of Cryptographic
  Techniques}, pages 186--194. Springer, 1986.

\bibitem{fouquet+morain02}
Mireille Fouquet and Fran\c{c}ois Morain.
\newblock Isogeny volcanoes and the {SEA} algorithm.
\newblock In Claus Fieker and David~R. Kohel, editors, {\em Algorithmic Number
  Theory Symposium}, volume 2369 of {\em Lecture Notes in Computer Science},
  pages 47--62, Berlin, Heidelberg, 2002. Springer Berlin / Heidelberg.

\bibitem{galbraith99}
Steven~D. Galbraith.
\newblock Constructing isogenies between elliptic curves over finite fields.
\newblock {\em LMS Journal of Computation and Mathematics}, 2:118--138, 1999.

\bibitem{galbraith2012mathematics}
Steven~D Galbraith.
\newblock {\em Mathematics of public key cryptography}.
\newblock Cambridge University Press, 2012.
\newblock
  \url{https://www.math.auckland.ac.nz/~sgal018/crypto-book/crypto-book.html}.

\bibitem{GHS}
Steven~D. Galbraith, Florian Hess, and Nigel~P. Smart.
\newblock Extending the {GHS} {W}eil descent attack.
\newblock In {\em Advances in cryptology--{EUROCRYPT} 2002 ({A}msterdam)},
  volume 2332 of {\em Lecture Notes in Comput. Sci.}, pages 29--44. Springer,
  Berlin, 2002.

\bibitem{galbraithsecurity}
Steven~D. Galbraith, Christophe Petit, Barak Shani, and Yan~Bo Ti.
\newblock On the security of supersingular isogeny cryptosystems.
\newblock In {\em Advances in Cryptology--ASIACRYPT 2016: 22nd International
  Conference on the Theory and Application of Cryptology and Information
  Security, Hanoi, Vietnam, December 4-8, 2016, Proceedings, Part I 22}, pages
  63--91. Springer, 2016.

\bibitem{cryptoeprint:2016:1154}
Steven~D. Galbraith, Christophe Petit, and Javier Silva.
\newblock Signature schemes based on supersingular isogeny problems.
\newblock Cryptology ePrint Archive, Report 2016/1154, 2016.
\newblock \url{http://eprint.iacr.org/2016/1154}.

\bibitem{gaudry+hess+smart02}
Pierrick Gaudry, Florian Hess, and Niegel Smart.
\newblock Constructive and destructive facets of {W}eil descent on elliptic
  curves.
\newblock {\em Journal of Cryptology}, 15(1):19--46--46, March 2002.

\bibitem{gelin2017loop}
Alexandre G{\'e}lin and Benjamin Wesolowski.
\newblock Loop-abort faults on supersingular isogeny cryptosystems.
\newblock In {\em International Workshop on Post-Quantum Cryptography}, pages
  93--106. Springer, 2017.

\bibitem{ionica+joux13}
Sorina Ionica and Antoine Joux.
\newblock Pairing the volcano.
\newblock {\em Mathematics of Computation}, 82(281):581--603, 2013.

\bibitem{jao+defeo2011}
David Jao and Luca De~Feo.
\newblock Towards {Quantum-Resistant} cryptosystems from supersingular elliptic
  curve isogenies.
\newblock In Bo-Yin Yang, editor, {\em Post-Quantum Cryptography}, volume 7071
  of {\em Lecture Notes in Computer Science}, pages 19--34, Berlin, Heidelberg,
  2011. Springer Berlin / Heidelberg.

\bibitem{JMV}
David Jao, Stephen~D. Miller, and Ramarathnam Venkatesan.
\newblock Expander graphs based on {GRH} with an application to elliptic curve
  cryptography.
\newblock {\em Journal of Number Theory}, 129(6), 2009.

\bibitem{jao2014isogeny}
David Jao and Vladimir Soukharev.
\newblock Isogeny-based quantum-resistant undeniable signatures.
\newblock In {\em International Workshop on Post-Quantum Cryptography}, pages
  160--179. Springer, 2014.

\bibitem{joux2009algorithmic}
Antoine Joux.
\newblock {\em Algorithmic cryptanalysis}.
\newblock CRC Press, 2009.

\bibitem{vercauteren-sidh-fp}
Angshuman Karmakar, Sujoy~Sinha Roy, Frederik Vercauteren, and Ingrid
  Verbauwhede.
\newblock Efficient finite field multiplication for isogeny based post quantum
  cryptography.
\newblock {\em Proceedings of WAIFI 2016}, 2016.

\bibitem{koblitz87}
Neal Koblitz.
\newblock Elliptic curve cryptosystems.
\newblock {\em Mathematics of Computation}, 48(177):203--209, 1987.

\bibitem{kohel}
David Kohel.
\newblock {\em Endomorphism rings of elliptic curves over finite fields}.
\newblock PhD thesis, University of California at Berkley, 1996.

\bibitem{kohel2014quaternion}
David Kohel, Kristin Lauter, Christophe Petit, and Jean-Pierre Tignol.
\newblock On the quaternion-isogeny path problem.
\newblock {\em LMS Journal of Computation and Mathematics}, 17(A):418--432,
  2014.

\bibitem{Kup}
Greg Kuperberg.
\newblock A subexponential-time quantum algorithm for the dihedral hidden
  subgroup problem.
\newblock {\em SIAM Journal of Computing}, 35(1):170--188, 2005.

\bibitem{lang1987elliptic}
Serge Lang.
\newblock {\em Elliptic Functions}, volume 112 of {\em Graduate texts in
  mathematics}.
\newblock Springer, 1987.

\bibitem{lenstra87}
Hendrik~W. Lenstra.
\newblock Factoring integers with elliptic curves.
\newblock {\em Annals of Mathematics}, 126:649--673, 1987.

\bibitem{lercier-algorithmique}
Reynald Lercier.
\newblock {\em Algorithmique des courbes elliptiques dans les corps finis}.
\newblock PhD thesis, LIX - CNRS, June 1997.

\bibitem{lercier+sirvent08}
Reynald Lercier and Thomas Sirvent.
\newblock On {E}lkies subgroups of \(\ell\)-torsion points in elliptic curves
  defined over a finite field.
\newblock {\em Journal de th\'{e}orie des nombres de Bordeaux}, 20(3):783--797,
  2008.

\bibitem{MOV}
Alfred Menezes, Scott Vanstone, and Tatsuaki Okamoto.
\newblock Reducing elliptic curve logarithms to logarithms in a finite field.
\newblock In {\em STOC '91: Proceedings of the twenty-third annual ACM
  symposium on Theory of computing}, pages 80--89, New York, NY, USA, 1991.
  ACM.

\bibitem{Mestre}
Jean-Fran\c{c}ois Mestre.
\newblock La m\'{e}thode des graphes. {E}xemples et applications.
\newblock In {\em Proceedings of the international conference on class numbers
  and fundamental units of algebraic number fields ({K}atata, 1986)}, Nagoya,
  1986. Nagoya University.

\bibitem{miller86}
Victor~S. Miller.
\newblock Use of elliptic curves in cryptography.
\newblock In {\em Lecture notes in computer sciences; 218 on Advances in
  cryptology--CRYPTO 85}, pages 417--426, New York, NY, USA, 1986.
  Springer-Verlag New York, Inc.

\bibitem{MiretMRV05}
Josep~M. Miret, Ramiro Moreno, Ana Rio, and Magda Valls.
\newblock Determining the 2-{S}ylow subgroup of an elliptic curve over a finite
  field.
\newblock {\em Mathematics of Computation}, 74(249):411--427, 2005.

\bibitem{MiretMSTV06}
Josep~M. Miret, Ramiro Moreno, Daniel Sadornil, Juan Tena, and Magda Valls.
\newblock An algorithm to compute volcanoes of 2-isogenies of elliptic curves
  over finite fields.
\newblock {\em Applied Mathematics and Computation}, 176(2):739--750, 2006.

\bibitem{cryptoeprint:2017:571}
Christophe Petit.
\newblock Faster algorithms for isogeny problems using torsion point images.
\newblock Cryptology ePrint Archive, Report 2017/571, 2017.
\newblock \url{http://eprint.iacr.org/2017/571}.

\bibitem{cryptoeprint:2017:962}
Christophe Petit and Kristin Lauter.
\newblock Hard and easy problems for supersingular isogeny graphs.
\newblock Cryptology ePrint Archive, Report 2017/962, 2017.
\newblock \url{http://eprint.iacr.org/2017/962}.

\bibitem{quis}
Christophe Petit, Kristin Lauter, and Jean-Jacques Quisquater.
\newblock Full cryptanalysis of {LPS} and {M}orgenstern hash functions.
\newblock In {\em Proceedings of the 6th international conference on Security
  and Cryptography for Networks}, SCN '08, Berlin, Heidelberg, 2008.
  Springer-Verlag.

\bibitem{pizer1}
Arnold~K. Pizer.
\newblock Ramanujan graphs and {H}ecke operators.
\newblock {\em Bulletin of the American Mathematical Society (N.S.)}, 23(1),
  1990.

\bibitem{pizer2}
Arnold~K. Pizer.
\newblock Ramanujan graphs.
\newblock In {\em Computational perspectives on number theory ({C}hicago, {IL},
  1995)}, volume~7 of {\em AMS/IP Stud. Adv. Math.} Amer. Math. Soc.,
  Providence, RI, 1998.

\bibitem{regev04}
Oded Regev.
\newblock A subexponential time algorithm for the dihedral hidden subgroup
  problem with polynomial space.
\newblock arXiv:quant-ph/0406151, June 2004.
\newblock \url{http://arxiv.org/abs/quant-ph/0406151}.

\bibitem{R&S}
Alexander Rostovtsev and Anton Stolbunov.
\newblock Public-key cryptosystem based on isogenies.
\newblock Cryptology ePrint Archive, Report 2006/145, 2006.
\newblock \url{http://eprint.iacr.org/2006/145}.

\bibitem{schoof85}
Ren\'{e} Schoof.
\newblock Elliptic curves over finite fields and the computation of square
  roots mod \(p\).
\newblock {\em Mathematics of Computation}, 44(170):483--494, 1985.

\bibitem{schoof95}
Ren\'{e} Schoof.
\newblock Counting points on elliptic curves over finite fields.
\newblock {\em Journal de Th\'{e}orie des Nombres de Bordeaux}, 7(1):219--254,
  1995.

\bibitem{shor1994algorithms}
Peter~W Shor.
\newblock Algorithms for quantum computation: Discrete logarithms and
  factoring.
\newblock In {\em Foundations of Computer Science, 1994 Proceedings., 35th
  Annual Symposium on}, pages 124--134. IEEE, 1994.

\bibitem{silverman:elliptic}
Joseph~H. Silverman.
\newblock {\em The arithmetic of elliptic curves}, volume 106 of {\em Graduate
  Texts in Mathematics}.
\newblock Springer-Verlag, New York, 1992.

\bibitem{silverman:advanced}
Joseph~H. Silverman.
\newblock {\em Advanced Topics in the Arithmetic of Elliptic Curves}, volume
  151 of {\em Graduate Texts in Mathematics}.
\newblock Springer, January 1994.

\bibitem{Stol}
Anton Stolbunov.
\newblock Constructing public-key cryptographic schemes based on class group
  action on a set of isogenous elliptic curves.
\newblock {\em Adv. Math. Commun.}, 4(2), 2010.

\bibitem{sun2012toward}
Xi~Sun, Haibo Tian, and Yumin Wang.
\newblock Toward quantum-resistant strong designated verifier signature from
  isogenies.
\newblock In {\em 2012 Fourth International Conference on Intelligent
  Networking and Collaborative Systems}, 2012.

\bibitem{sutherland10}
Andrew~V. Sutherland.
\newblock Genus 1 point counting over prime fields.
\newblock Last accessed July 16, 2010.
  \url{http://www-math.mit.edu/\~drew/SEArecords.html}, 2010.

\bibitem{tani2009claw}
Seiichiro Tani.
\newblock Claw finding algorithms using quantum walk.
\newblock {\em Theoretical Computer Science}, 410(50):5285--5297, 2009.

\bibitem{tao2011expander}
Terence Tao.
\newblock Expansion in groups of {L}ie type -- basic theory of expander graphs.
\newblock
  \url{https://terrytao.wordpress.com/2011/12/02/245b-notes-1-basic-theory-of-expander-graphs/},
  2011.

\bibitem{teske06}
Edlyn Teske.
\newblock An elliptic curve trapdoor system.
\newblock {\em Journal of Cryptology}, 19(1):115--133, January 2006.

\bibitem{ti2017fault}
Yan~Bo Ti.
\newblock Fault attack on supersingular isogeny cryptosystems.
\newblock In {\em International Workshop on Post-Quantum Cryptography}, pages
  107--122. Springer, 2017.

\bibitem{tillich2008collisions}
Jean-Pierre Tillich and Gilles Z{\'e}mor.
\newblock Collisions for the lps expander graph hash function.
\newblock In {\em Annual International Conference on the Theory and
  Applications of Cryptographic Techniques}, pages 254--269. Springer, 2008.

\bibitem{velu71}
Jean V{\'{e}}lu.
\newblock Isog{\'{e}}nies entre courbes elliptiques.
\newblock {\em Comptes Rendus de l'Acad\'{e}mie des Sciences de Paris},
  273:238--241, 1971.

\bibitem{waterhouse69}
William~C. Waterhouse.
\newblock Abelian varieties over finite fields.
\newblock {\em Annales Scientifiques de l'\'{E}cole Normale Sup\'{e}rieure},
  2(4):521--560, 1969.

\bibitem{cryptoeprint:2017:186}
Youngho Yoo, Reza Azarderakhsh, Amir Jalali, David Jao, and Vladimir Soukharev.
\newblock A post-quantum digital signature scheme based on supersingular
  isogenies.
\newblock Cryptology ePrint Archive, Report 2017/186, 2017.
\newblock \url{http://eprint.iacr.org/2017/186}.

\end{thebibliography}

\end{document}